%% file: main.tex
\documentclass[sigconf,preprint]{acmart}
\AtBeginDocument{%
  }

\setcopyright{acmlicensed}
\copyrightyear{2018}
\acmYear{2018}
\acmDOI{XXXXXXX.XXXXXXX}
\acmConference[Conference acronym 'XX]{Make sure to enter the correct
  conference title from your rights confirmation email}{June 03--05,
  2018}{Woodstock, NY}
\acmISBN{978-1-4503-XXXX-X/2018/06}

\usepackage{subcaption}
\usepackage{svg}
\usepackage[table,xcdraw]{xcolor}
\usepackage{algorithm}
\usepackage{algpseudocode}
\usepackage{multirow}

\newcommand{\trim}{\vspace{-2mm}}
\newcommand{\stitle}[1]{\vspace*{0.4em}\noindent{\bf #1.\/}}
\newcommand{\squishlist}{
	\begin{list}{$\bullet$}
		{ \setlength{\itemsep}{1pt}
			\setlength{\parsep}{1pt}
			\setlength{\topsep}{2.5pt}
			\setlength{\partopsep}{0.5pt}
			\setlength{\leftmargin}{1em}
			\setlength{\labelwidth}{1em}
			\setlength{\labelsep}{0.6em}
		}
	}
	\newcommand{\squishend}{
	\end{list}
}

\newcommand{\circnum}[1]{\textcircled{\scriptsize #1}}

\begin{document}

\title{FuxiShuffle: An Adaptive and Resilient Shuffle Service for Distributed Data Processing on Alibaba Cloud}

\author{Yuhao Lin$^{\dagger}$, Zhipeng Tang$^{\star}$, Jiayan Tong$^{\star}$, Junqing Xiao$^{\star}$, Bin Lu$^{\star}$, Yuhang Li$^{\star}$, Chao Li$^{\star}$\\ Zhiguo Zhang$^{\star}$, Junhua Wang$^{\star}$, Hao Luo$^{\dagger}$, James Cheng$^{\ddagger}$, Chuang  Hu$^{\dagger}$, Jiawei Jiang$^{\dagger}$, Xiao Yan$^{\dagger}$
}
\affiliation{
  \institution{$^{\dagger}$Wuhan University, $^{\star}$Alibaba Cloud, Alibaba Group, $^{\ddagger}$The Chinese University of Hong Kong}
  \city{}
  \country{}
  }
\email{{yuhao\_lin, lohozz, handc, jiawei.jiang, yanxiaosunny}@whu.edu.cn, jcheng@cse.cuhk.edu.hk}
\email{{zhipeng.tzp, tongjiayan.tjy, junqing.xjq, bin.lub, yuzhou.lyh, li.chao, zhiguo.z, junhua.w}@alibaba-inc.com}



\renewcommand{\shortauthors}{Lin et al.}

\input{chapters/0-abstract}



\maketitle

\input{chapters/1-Introduction}

\input{chapters/2-Motivation}

\input{chapters/3-System}

\input{chapters/4-Adaptability}

\input{chapters/5-Resilience}

\input{chapters/6-Evaluation}
\input{chapters/7-Lessons_in_Practice}

\input{chapters/8-Related_Work}
\input{chapters/9-Conclusion}

\bibliographystyle{ACM-Reference-Format}
\bibliography{main}










\end{document}

%% file: chapters/0-abstract.tex

\begin{abstract}
Shuffle exchanges intermediate results between upstream and downstream operators in distributed data processing and is usually the bottleneck due to factors such as small random I/Os and network contention. Several systems have been designed to improve shuffle efficiency, but from our experiences of running ultra-large clusters at Alibaba Cloud MaxCompute platform, we observe that they can not adapt to highly dynamic job characteristics and cluster resource conditions, and their fault tolerance mechanisms are passive and inefficient when failures are inevitable. To tackle their limitations, we design and implement FuxiShuffle as a general data shuffle service for the ultra-large production environment of MaxCompute, featuring good adaptability and efficient failure resilience. Specifically, to achieve good adaptability, FuxiShuffle dynamically selects the shuffle mode based on runtime information, conducts progress-aware scheduling for the downstream workers, and automatically determines the most suitable backup strategy for each shuffle data chunk. To make failure resilience efficient, FuxiShuffle actively ensures data availability with multi-replica failover, prevents memory overflow with careful memory management, and employs an incremental recovery mechanism that does not lose computation progress. Our experiments show that, compared to baseline systems, FuxiShuffle significantly reduces not only end-to-end job completion time but also aggregate resource consumption. Micro experiments suggest that our designs are effective in improving adaptability and failure resilience.  
\end{abstract}

%% file: chapters/1-Introduction.tex
\section{Introduction}\label{sec:intro}

\begin{figure}[!t]
  \centering
  \includegraphics[width=0.8\linewidth]{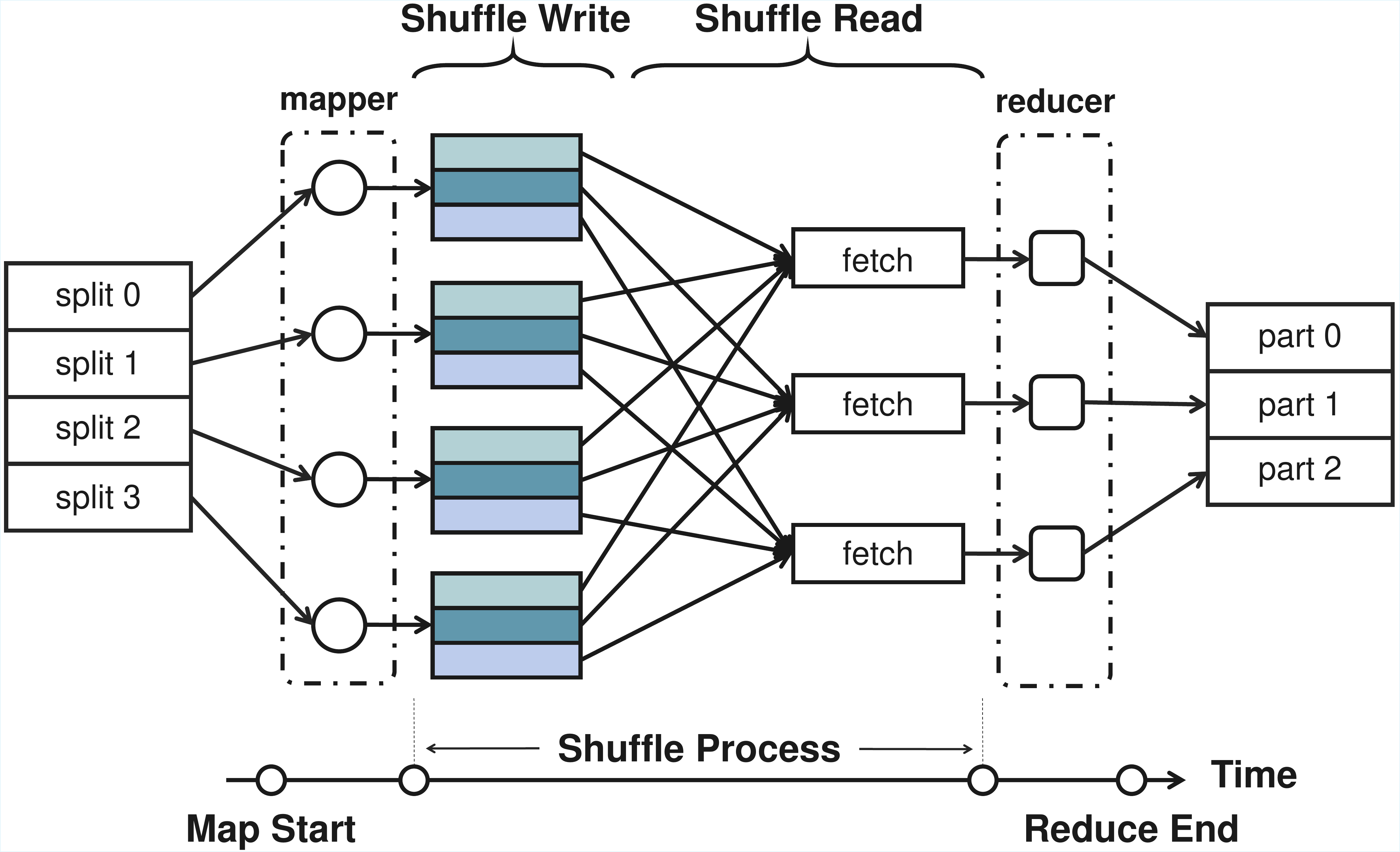}
  \trim
  \caption{An illustration of the shuffle process.}
  \trim
  \label{fig:mapreduce shuffle}
\end{figure}

In industry, data processing often operates at petabyte (PB) or exabyte (EB) scale for tasks such as analyzing the statistics of online applications (e.g., e-commerce and video streaming), mining user or item patterns (e.g., for recommendation or business planning), and preparing data for model training~\cite{jiang2020alibaba, zhou2023ahpa, gatlin2024real, moina2024cloud, greca2020impact, nabli2022description, zhang2025streaming, chen2025scheduling, jiang2021towards}. To execute such large-scale data analytics, many distributed data processing systems are developed, including Apache Hadoop~\cite{apache_hadoop}, Apache Spark~\cite{apache_spark}, Google’s FlumeJava \cite{chambers2010flumejava}, Cloud Dataflow \cite{krishnan2015google}, and Microsoft’s Cosmos/Scope \cite{patel2019big, chaiken2008scope}. These systems typically offer a rich set of operators for users to compose data-processing jobs, where each job models operator dependencies as a directed acyclic graph (DAG)~\cite{tang2020survey}. 
For parallel execution, multiple instances are spawned for each operator (e.g., table scan, merge join, or sort) to process data partitions distributed across servers.
On Alibaba Cloud, we run the MaxCompute platform~\cite{wang2018billion, weng2018online} to serve distributed data processing jobs for both our internal users and public cloud customers. 
Currently, MaxCompute handles about several EB of data on a daily basis, and MaxCompute clusters can comprise over 100K servers. 
Due to the immense scale, 
it is crucial for MaxCompute to improve execution efficiency and stability, ensuring good user experience and low operating costs.


As a core phase in distributed data processing, shuffle exchanges data among different operators~\cite{daikoku2016exploring, chen2012joint, fan2020shuffle}. Figure~\ref{fig:mapreduce shuffle} provides an example within the MapReduce framework: mappers generate intermediate results for their local data shards, and these results are transferred to reducers for further aggregation or computation~\cite{wang2015phase, ahmad2014shufflewatcher, yu2013virtual}. More generally, shuffle moves data from an upstream operator (called \textit{writer}) to a downstream operator (called \textit{reader}), and this process can involve partitioning, sorting, and merging of data blocks. Shuffle often becomes a major performance bottleneck in distributed data processing systems~\cite{zhang2018riffle, shen2020magnet, guo2016ishuffle, liu2017optimizing, liang2016bashuffler, guo2014exploiting}, often accounting for more than 50\% of the end-to-end job execution time~\cite{li2015sparkbench, fu2018efficient, wang2015phase, nicolae2016leveraging}. This is because shuffle process typically traverses the network, potentially across multiple racks or data centers, and incurs substantial overheads including random disk I/O for small data blocks, serialization and deserialization costs, network contention under concurrent operator execution. 


Several systems aim to improve shuffle performance~\cite{liu2017optimizing, rao2012sailfish, wang2011hadoop}. 
For instance, HDShuffle~\cite{qiao2019hyper} and Hadoop-A~\cite{wang2011hadoop} merge intermediate results to reduce fragmented shuffle data. iShuffle~\cite{guo2016ishuffle} decouples data writing from the upstream and downstream operators to accelerate the shuffle process. SCache \cite{fu2018efficient} uses size prediction with in-memory caching for pre-scheduling and prefetching. Magnet \cite{shen2020magnet} pushes fragmented blocks to remote services for partition-level merging and improved locality.
Given that failures (e.g., OOM, server or network issues) are unavoidable in distributed systems, existing approaches provide fallback or retry mechanisms. 
Riffle~\cite{zhang2018riffle} and Magnet~\cite{shen2020magnet} mitigates I/O bottlenecks by merging shuffle files while retaining originals, and fall back upon merge failures; Sailfish~\cite{rao2012sailfish} uses a block-level index to locate lost data and re-executes only the affected upstream tasks to rebuild new version data; 
and OPS~\cite{cheng2020ops} also re-executes only the tasks on the failed nodes, avoiding large-scale recomputation due to data loss.


\stitle{Limitations of Existing Solutions} When optimizing shuffle performance for MaxCompute, we find that existing techniques are limited in two crucial aspects due to the large cluster scale, high job concurrency, and complex resource dynamics in our production:

\squishlist

\item \textit{Poor adaptability.} Existing systems use a fixed shuffle mode (i.e., memory or disk) for all jobs \cite{apache_hadoop, zhang2018riffle, guo2016ishuffle, rao2012sailfish}, or spill to disk when memory is insufficient at best, and thus they cannot choose the shuffle mode according to runtime resource conditions and per-job characteristics.
For shuffle data reading, they adopt either \textit{staged scheduling} \cite{dean2008mapreduce}, which starts the readers after all writers completes and can cause a long waiting time due to straggled writers, or \textit{gang scheduling}~\cite{wiseman2003paired}, which starts the writers and readers simultaneously and can waste resources when the readers are left idle waiting for data. Moreover, existing systems use a fixed storage layout for the shuffle data and do not account for factors such as varying data sizes and skewness, server-side resource availability, and service level agreement (SLA) of jobs.

\item \textit{Inefficient fault tolerance.} 
As computing clusters scale out and the number of tasks increases, failures become common \cite{li2020predicting, vishwanath2010characterizing, mohammed2017failover, yang2016reliable}. However, existing systems largely rely on passive fault-tolerance mechanisms such as data replication and task re-execution, which are inefficient for frequent failures. In particular, these systems may create replicas for all shuffle data, which ensures robustness but introduces substantial overheads. Moreover, when a downstream task encounters read errors, they typically recover by re-executing the upstream tasks. This forces the downstream tasks to wait for the upstream and causes all previously fetched data and computation progress to be discarded, which may even trigger repeated re-executions \cite{zhang2014fuxi} and lead to amplified job latencies or timeouts.

\squishend

\stitle{FuxiShuffle} By tackling the limitations of existing systems regarding adaptability and fault tolerance, we design and implement FuxiShuffle as the shuffle service for MaxCompute. 
FuxiShuffle is general in that it can serve data shuffling for all shuffle-inducing operators (e.g., join, repartition, aggregate, and reduce), and has been integrated with multiple distributed data processing systems—including Alibaba MaxCompute and Spark on Maxcompute. 
FuxiShuffle adopts an architecture of managers and workers to decompose data and control planes. The Shuffle Agents reside on both the compute and storage servers and are responsible for collecting, aggregating, and storing the shuffle data, while the Job Manager in Fuxi \cite{zhang2014fuxi, chen2021fangorn} and the Shuffle Service Manager conduct workers and Shuffle Agents scheduling in the shuffle process.

To handle dynamic resource conditions and job characteristics, FuxiShuffle introduces adaptive mechanisms across the entire shuffle life cycle. In particular, to select the optimal shuffle mode (i.e., in-memory or on-disk), we utilize an execution time threshold to find jobs that benefit the most from in-memory shuffle and adjust the threshold dynamically according to idle memory resources. 
To schedule workers during shuffle data reading, we introduce \textit{progressive scheduling} to actively schedule the downstream readers and retrieve data from the writers in a streaming manner based on execution progress and resource status. Moreover, for each partition of the shuffle data, FuxiShuffle dynamically determines whether to perform aggregated writing (for performance) or replicated backup (for reliability) to balance between efficiency and robustness. 

To enhance fault tolerance, FuxiShuffle introduces proactive resilience mechanisms. 
Specifically, FuxiShuffle groups the writers by destination data partition and assigns each group to a Shuffle Agent, forming \textit{Shuffle Agent Groups} to mitigate the multi-to-one network pressure. Each Shuffle Agent also has replicas to allow dynamic failover in the case of single point of failures.
FuxiShuffle also conducts careful memory management on the servers and adopts a priority-based eviction strategy when memory is insufficient to avoid out-of-memory (OOM) errors. 
To recover efficiently from failures, FuxiShuffle supports \textit{incremental recovery}, which allows readers to continue computing while writers are re-executed, instead of interrupting the readers and restarting from scratch.

We comprehensively evaluate FuxiShuffle on the production clusters of Alibaba Cloud. The results show that it substantially outperforms the baseline systems, reducing end-to-end job completion time by 76.36\% on average and cutting the aggregate resource consumption by 67.14\%. In the case of single point of failures, the performance degradation remains within 10\%, and under continuous disturbances, the system still maintains efficiency and robustness. Currently,  FuxiShuffle stably processes several EBs of shuffle data per day, involving tens of billions of worker interactions and executing tens million of DAG jobs, while maintaining excellent performance and reliability. To guide followup works, we also summarize the insights for deploying shuffle service in large-scale clusters.

\begin{figure*}[!t]
  \centering
  \begin{minipage}[!t]{0.32\textwidth}
    \centering
    \includegraphics[width=0.75\columnwidth]{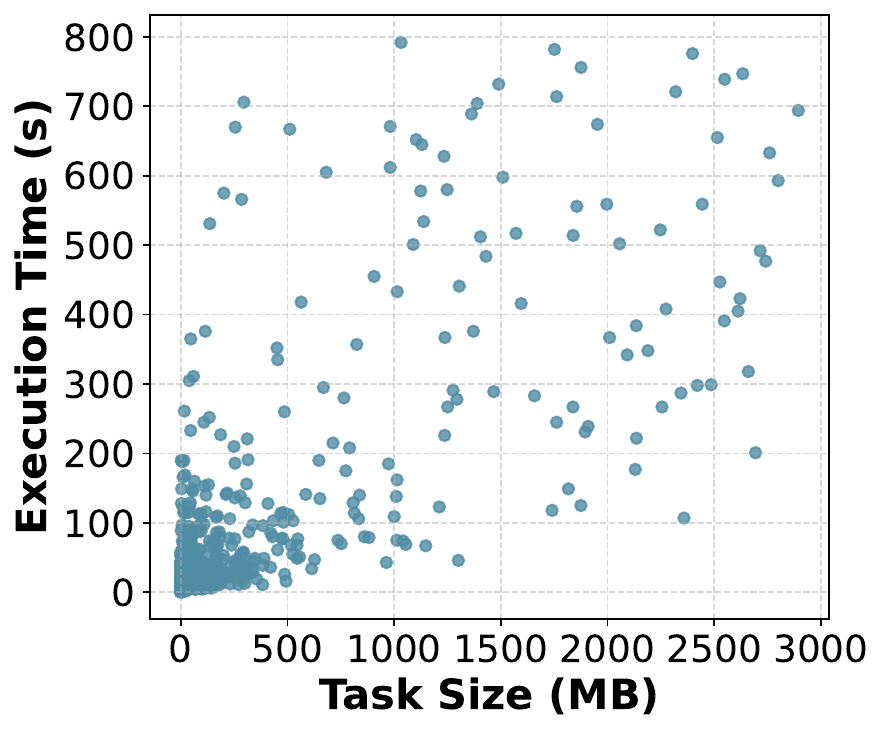}
    \trim
    \caption{The distributions for the size (measured by input data) and execution time of the tasks in our cluster. }
    \trim
    \label{fig:shuffle_mode}
  \end{minipage}
  \hfill
  \begin{minipage}[!t]{0.32\textwidth}
    \centering
    \includegraphics[width=0.75\columnwidth]{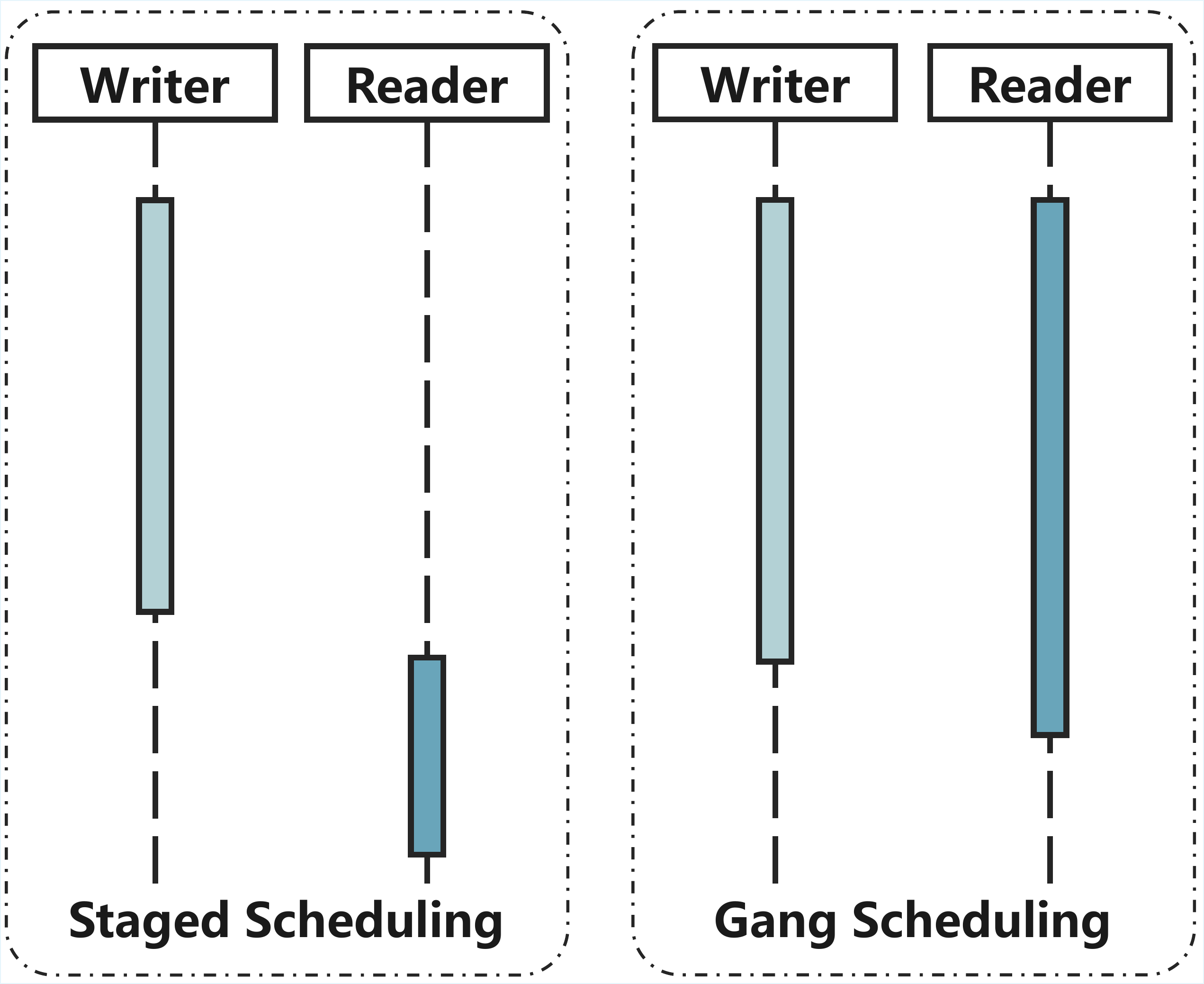}
    \trim
    \caption{The task launch patterns of the staged scheduling and gang scheduling strategies in existing systems.}
    \trim
    \label{fig:launch_patterns}
  \end{minipage}
  \hfill
  \begin{minipage}[!t]{0.32\textwidth}
    \centering
    \includegraphics[width=0.75\linewidth]{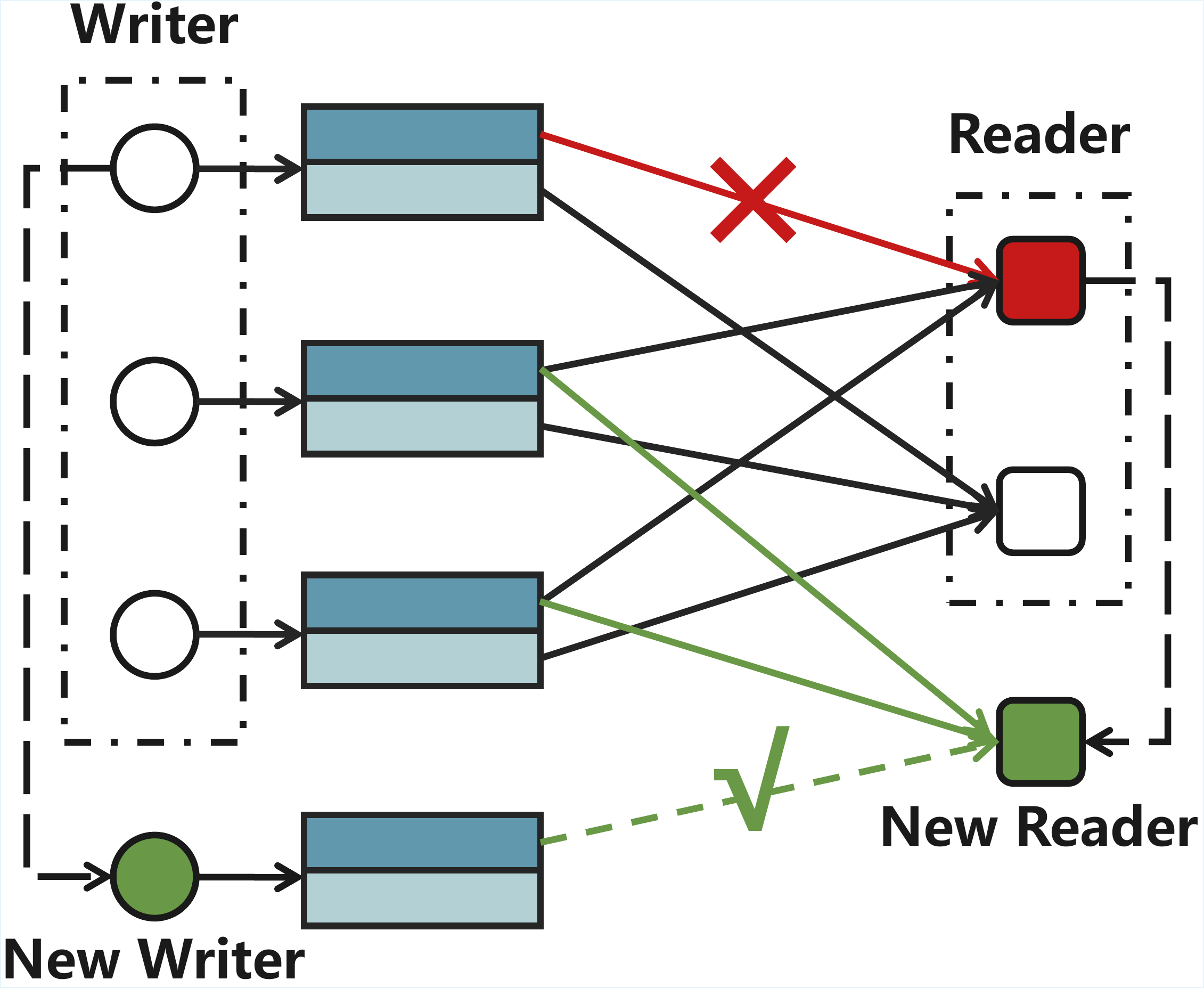}
    \trim
     \caption{Partial re-execution. A failed writer is re-executed, the reader waits for the writer and is also re-executed.}
    \trim
    \label{fig:re-execution}
  \end{minipage}
\end{figure*}

Overall, we make the following contributions:
\squishlist

\item Guided by the experiences from running large-scale production clusters, we analyze two critical limitations of existing data shuffle systems for distributed data processing, i.e., poor adaptability and inefficient failure recover. (Section~\ref{sec:motivation}) 

\item We present the architecture, deployment and workflow of FuxiShuffle as a general and performant shuffle service in large-scale production environments. (Section~\ref{sec:overall})

\item We adaptively manage shuffle mode selection, shuffle data reading, and data layout planning to cope with dynamic cluster conditions and job characteristics. (Section~\ref{sec:adaptive})

\item We provocatively prevent failures using replica redundancy and fine-grained memory management and improve the efficiency of failure recovery with incremental recovery.  (Section~\ref{sec:recovery})

\squishend

%% file: chapters/2-Motivation.tex
\section{Motivations}
\label{sec:motivation}

In this part, we analyze the challenges of deploying shuffle service in large-scale clusters based on our experiences. 

\stitle{Challenges in Adaptability} Existing shuffle systems usually adopt a fixed shuffle mode, data reading strategy, and data layout policy for all jobs. Such a one-size-fits-all design cannot adapt to the dynamic variations in data distributions, resource conditions, and service level agreement (SLA) for jobs in large clusters.

\squishlist

\item \textit{Fixed shuffle mode.} In-memory shuffle offers superior performance than on-disk shuffle, but it is infeasible to utilize in-memory shuffle for all tasks due to limited memory resources. As shown in Figure~\ref{fig:shuffle_mode}, the data size and execution time of the tasks exhibit large variances. 
Tasks with smaller data size and shorter execution time are more suitable for in-memory shuffle because they will occupy smaller memory for a shorter time. 
Moreover, in-memory shuffle should be used more aggressively when cluster memory is abundant but more conservatively when memory pressure is high.
However, existing systems either use a fixed shuffle mode for all jobs or require the user to manually configure the shuffle model for each job.

\item \textit{Inefficient data reading.} Figure \ref{fig:launch_patterns} illustrates the staged scheduling and gang scheduling \cite{wiseman2003paired} in an early version of Alibaba Cloud’s Fuxi system \cite{zhang2014fuxi, chen2021fangorn}. In particular, staged scheduling only starts the downstream readers after all upstream writers finish, leading to long end-to-end latency; gang scheduling starts all upstream and downstream workers simultaneously, enabling faster execution but will degrade resource utilization if the downstream readers spend a long idle time waiting for data. Overall, staged scheduling and gang scheduling show the tension between job latency and resource utilization, and new data reading methods are required to achieve a good trade-off between the two aspects.

\item \textit{Rigid data layout.} Most existing systems use static data layout policies that cannot adapt to skewed or fluctuating workloads. Hadoop lacks intermediate aggregation, causing small fragmented files to create I/O bottlenecks. 
Some systems (e.g., OPS \cite{cheng2020ops}, Sailfish \cite{rao2012sailfish}) introduce data aggregation but fix the aggregation granularity at the reader partition level. When a single partition is large (e.g., 10 GB), these systems cannot bypass the aggregation path to causing resource pressure. They also cannot dynamically enable reliable backups for the critical partitions, often triggering task retries or even job failures. Moreover, although full local backups improve fault tolerance (as in Riffle \cite{zhang2018riffle} and Magnet \cite{shen2020magnet}), they compete with normal shuffle traffic for disk bandwidth under high concurrency, usually degrading overall performance.


\squishend

\begin{figure*}[!t]
  \centering
  \includegraphics[width=\textwidth]{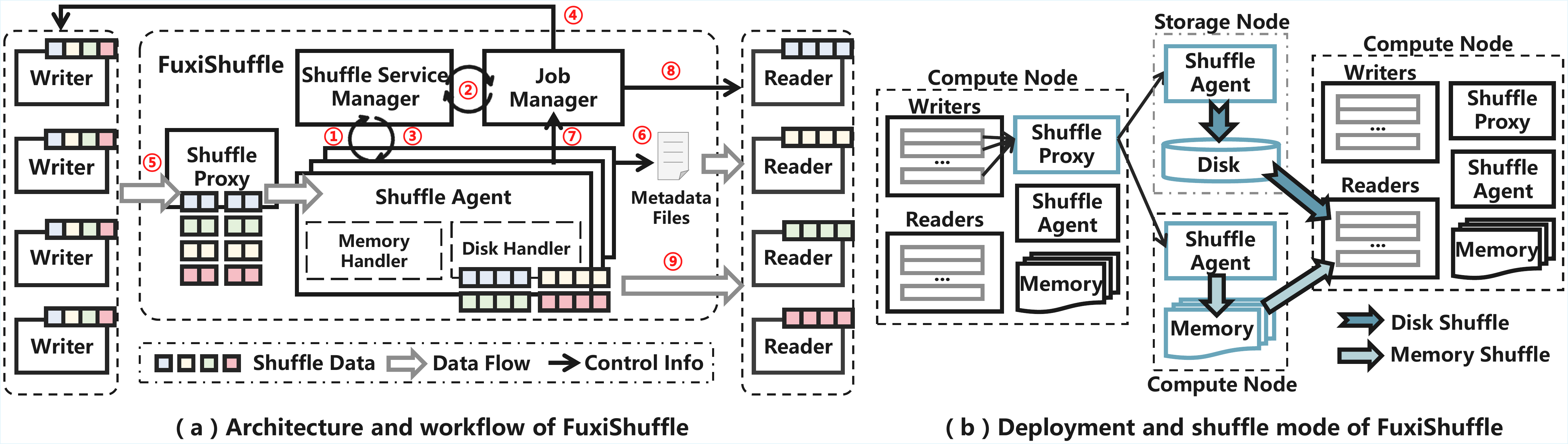}
  \trim
  \caption{The architecture and workflow of FuxiShuffle (left plot), the deployment of FuxiShuffle (right plot).}
  \trim
  \label{fig:architecture_deployment}
\end{figure*}

\stitle{Challenges in Fault Tolerance}
In large-scale computing clusters, failures from hardware faults, network jitter, or resource contention~\cite{cotroneo2019bad, rouholamini2024proactive, luo2019improving, kirti2024fault} often prevent shuffle data from reaching downstream tasks, and existing systems typically rely on passive fault-tolerance mechanisms, which frequently result in prolonged shuffle completion times or even job failures.

\squishlist

\item \textit{Single point of failure.} 
To address the low disk I/O efficiency and network congestion caused by fragmented reads of shuffle data, some existing systems introduce a centralized component between the writers and readers to collect and consolidate the shuffle data (i.e., Shuffle Agent in our case). As the scale of clusters and jobs grow, the number of such central components increases, and thus failures are more likely to affect them. As such, considerations should be made to account for the single point of failure in these central components, and the fault tolerance mechanism should make its additional overhead low.

\item \textit{Passive failure handling.} 
The shuffle processes of most systems (such as Hadoop and Spark) rely on passive failure recovery. That is, when an error occurs, the systems forcibly terminate the downstream tasks and re-execute the upstream stages, as illustrated in Figure~\ref{fig:re-execution}. 
To reduce the frequency of such failures and thereby lower the overhead of failure recovery, mechanisms are required to ensure shuffle data availability. An effective mechanism should actively prevent failures and ensure that downstream readers can reliably access shuffle data even in the presence of transient faults such as memory overflows or node failures.

\item \textit{Inefficient re-execution.} The re-execution in Figure~\ref{fig:re-execution} requires the downstream reader to wait for the upstream writer while the writer reconstructs lost data. Moreover, the computation progresses of the affected readers are discarded to start the computation from scratch. This prolongs the job execution time and may cause jobs to fall into \textit{repeated re-execution loops}.
The repeated re-execution loops can amplify the completion time of a job by several to tens of times, and may even cause the job to exceed its deadline and fail.
Our statistics show that about 11k jobs experience at least one partial re-execution on a daily basis, and 36\% of these jobs undergo more than 10 re-executions. 
As such, to reduce the impact of failures, re-execution should be made more efficient by avoiding waiting and loosing progress.

\squishend

%% file: chapters/3-System.tex
\section{FuxiShuffle Overview}\label{sec:overall}

In this section, we provide an overview for FuxiShuffle in terms of system architecture, shuffle workflow, and metadata management. The core designs of FuxiShuffle for adaptability and fault tolerance will be discussed in subsequent sections. 

\subsection{System Architecture}


\stitle{System Roles} Figure~\ref{fig:architecture_deployment} (a) shows the  overall architecture of FuxiShuffle, which consists of four core system roles that interact though a manager–worker structure that decouples control logic from data processing. We introduce the roles as follows:

\squishlist

\item \textit{Shuffle Agent} is a data-plane worker deployed on each machine. It receives partitioned data from upstream tasks, performs local aggregation, and writes shuffle data to memory or disk based on the selected shuffle mode (cf. Section~\ref{subsec:selection}). By merging fragmented outputs into contiguous, partition-aligned blocks, it enables efficient data reading for the downstream tasks.

\item \textit{Shuffle Proxy} sits between the writers and Shuffle Agents to decouple data writing and mitigate network congestion. Co-located with the writers, it batches small packets and performs preliminary per-partition aggregation locally before forwarding the data to the Shuffle Agents to improve network throughput.

\item \textit{Job Manager} bridges the computation frameworks and shuffle service and acts as the job-level scheduler. It selects the shuffle mode (in-memory or on-disk) based on task priority, execution time, data size, and cluster resources, assigns Shuffle Agents for jobs, and coordinates when to launch downstream readers.

\item \textit{Shuffle Service Manager} serves as the global control plane and conducts overall resource management, allocation, and health monitoring for the Shuffle Agents.

\squishend

\stitle{Deployment and Shuffle Modes}
To support large-scale deployment and enhance flexibility, FuxiShuffle adopts \textit{storage-compute} disaggregation as shown in Figure \ref{fig:architecture_deployment} (b). The Shuffle Agents can reside on both the compute nodes and storage nodes. On the compute nodes, the Shuffle Agents co-locate with the computing workers, while on the storage nodes, only Shuffle Agents are deployed. 
FuxiShuffle can work in both in-memory mode and on-disk shuffle mode, and the decision is made by the Job Manager for each task.


\squishlist

\item \textit{In-memory shuffle} keeps the shuffle data in the shared memory of the Shuffle Agents, and readers can either fetch  data locally from co-locating Shuffle Agents or remotely via network.

\item \textit{On-disk shuffle} persists shuffle data on Alibaba Cloud Pangu’s \cite{huang2019yugong} disk storage, and readers access data via network.

\squishend

As a multi-tenant platform, each node in MaxCompute can run dozens to hundreds of concurrent job instances from different users. FuxiShuffle ensures job stability via shared memory isolation and priority-based scheduling (cf. Section~\ref{subsec:memory management}).

\subsection{Shuffle Workflow}
As shown in Figure \ref{fig:architecture_deployment} (a), FuxiShuffle takes three major steps for a shuffle process, which we detail as follows.


\stitle{Initialization}
To prepare for service, the Shuffle Agents periodically report their runtime status, such as memory usage and load, to the Shuffle Service Manager for governance (Step \circnum{1}).
Upon the start of a job, the Job Manager registers the job with the Shuffle Service Manager (Step \circnum{2}). 
Based on predefined allocation policies, the Shuffle Service Manager assigns a list of Shuffle Agents to the writer instances of the job. 
The Shuffle Service Manager also sends information of the authorized data groups to Shuffle Agents such that only the data of active jobs is retained (Step \circnum{3}).

\stitle{Data Writing}
The Job Manager determines the appropriate shuffle mode (in-memory or on-disk) and partitioning strategy based on factors including job data size and cluster load, and conveys this information to the writer instances (Step \circnum{4}). Writers send data to the Shuffle Proxy for aggregation, and the aggregated data is then forwarded to the Shuffle Agents (Step \circnum{5}). Each Shuffle Agent aggregates and stores its incoming shuffle data and generates files for metadata (Step \circnum{6}). This file enables precise data addressing for downstream consumers and supports version control.

\stitle{Data Reading}
The downstream tasks autonomously locate and consume shuffle data without centralized coordination. Leveraging the Pre-Start and Pre-Read mechanisms (cf. Section~\ref{sec:data_fetching}), the Job Manager intelligently selects the optimal timing to launch the reader tasks based on the progress of the writers (Step \circnum{7}, \circnum{8}). 
Readers then dynamically merge and de-duplicate multiple metadata information from different sources to construct the target read locations and commence data reading (Step \circnum{9}). 
In this process, if a shuffle data partition becomes unavailable or corrupted, the reader identifies the missing data segments using the metadata files, automatically switches to redundant data sources, or  triggers partial re-execution for the upstream tasks to reconstruct the lost data.

\subsection{Metadata Management}
\label{sec:metadata_management}


\stitle{Meta Files}
FuxiShuffle adopts a decentralized shuffle data discovery mechanism based on metadata files. In particular, FuxiShuffle leverages standardized naming conventions and structured metadata such that the readers can locate and read the required data on their own. 
FuxiShuffle maintains two types of meta files, i.e., \textit{Primary Index} and \textit{Backup Index}, to record the paths for shuffle data stored in the Shuffle Agent Files and Backup Files (cf. Section \ref{sec:data-aware_layout}).

\squishlist

\item \textit{Primary Index} is generated by the Shuffle Agents after data aggregation and uniquely identified by the quintuple \textsf{⟨job\_id, task\_id, access\_point, partition\_id, agent\_ip⟩}, indicating that the index belongs to a specific job and task, is produced by a designated logical write stream, and is collected and merged into the corresponding partition by the specified Shuffle Agent.
The index records, for each data block, its originating writer, version number (RetryIdx), backup sequence number, offset, and length. For version control, the reader selects the block with the largest RetryIdx to access the latest data.

\item \textit{Backup Index} is produced by each writer when writing its shuffle data to a Backup File (usually located on the disk). Organized by partitions, the Backup Index records the offset and length of each data block within the Backup File, enabling readers to quickly locate data upon Shuffle Agent failures.

\squishend

\stitle{Data Reading}
During the read phase, the reader autonomously constructs the shuffle data read paths based on the meta files written by the Shuffle Agent. It then parses and de-duplicates the shuffle data written by writers for the same partition in the order of RetryIdx, generating a complete list of shuffle data retrieval information, including file paths, offsets, and lengths. Compared with existing systems that rely on a centralized scheduler to explicitly assign data read paths~\cite{apache_hadoop, apache_spark}, our index-based mechanism decouples data production from consumption, providing higher flexibility and robust fault tolerance for large-scale clusters.


%% file: chapters/4-Adaptability.tex
\section{Adaptive Control and Scheduling}
\label{sec:adaptive}


FuxiShuffle overcomes the limited adaptability of existing shuffle systems with a unified framework for shuffle mode selection, data reading, and layout planning, allowing it to adapt to workloads and runtime conditions while improving performance and flexibility.

\begin{figure}[t]
  \centering
  \includegraphics[width=0.75\columnwidth]{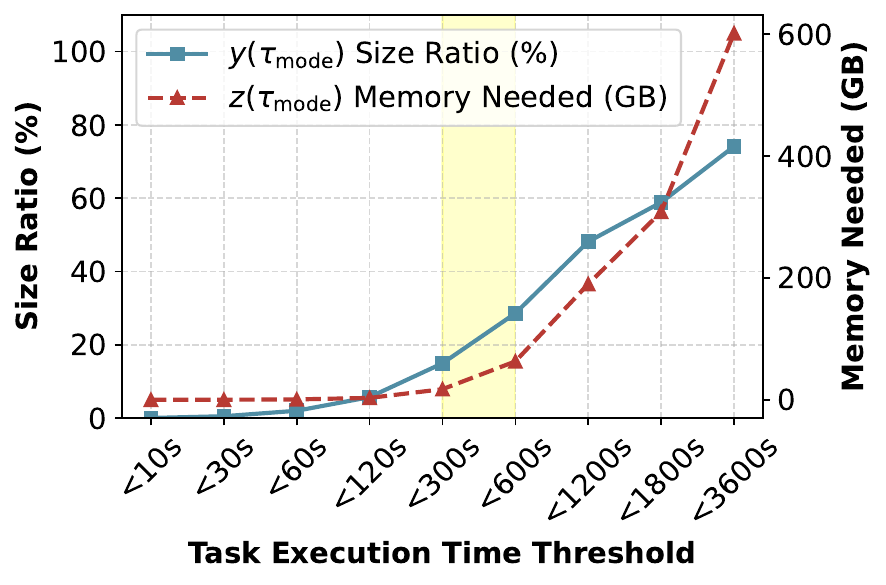}
  \trim
  \caption{The ratio of shuffle data that uses in-memory shuffle (left axis, measured by data size) and average memory consumption on each server (right axis) when adjusting the task execution time threshold (x axis) for in-memory shuffle.}
  \trim
  \label{fig:shuffle_mode_threshold}
\end{figure}

\subsection{Shuffle Mode Selection}
\label{subsec:selection}

To maximize the use of idle memory and accelerate more shuffle traffic, we designed an admission mechanism that adaptively determines whether a task is suitable for in-memory shuffle. The core idea is that tasks with small size and short execution time are suitable for placing shuffle data in memory to avoid disk I/O overhead, while long-running tasks or those with large partitions are better handled with on-disk shuffle to conserve memory.


The shuffle mode for each task is determined as follows:
\[
\text{Mode} = 
\begin{cases}
\text{In-memory Shuffle}, & \text{if } \hat{t} \leq \tau_{\text{mode}}, \\
\text{On-disk Shuffle},    & \text{otherwise}.
\end{cases}
\]
The core of this mechanism lies in dynamically selecting an execution time admission threshold $\tau_{\text{mode}}$: if the predicted runtime of a task $\hat{t} \leq \tau_{\text{mode}}$, in-memory shuffle is enabled; otherwise, on-disk shuffle is used.
Since the actual runtime of a task is unknown before scheduling, we estimate $\hat{t}$ by job attributes including input data volume, shuffle operator type (e.g., Join or GroupBy), key cardinality estimation, and historical task performance profiles. 

The selection of $\tau_{\text{mode}}$ must strike a balance between acceleration gains and memory consumption. To this end, we leverage historical task data to build prior knowledge and make decisions based on real-time resource status. Figure~\ref{fig:shuffle_mode_threshold} presents historical task log statistics from a production cluster, which includes two key functions:

\squishlist
\item $y(\tau_{\text{mode}})$: the ratio of shuffle data that can be accelerated by in-memory shuffle when the admission threshold is set to $\tau_{\text{mode}}$;
\item $z(\tau_{\text{mode}})$: the average memory needed per machine when the admission threshold is set to $\tau_{\text{mode}}$.
\squishend

Practical experience indicates that both $y(\tau_{\text{mode}})$ and $z(\tau_{\text{mode}})$ are non-decreasing functions of $\tau_{\text{mode}}$. Therefore, under the constraint of the current available memory upper bound $Z_{\text{available}}$, the optimal threshold $\tau_{\text{mode}}^*$ can be formalized as the following constrained optimization problem:
\[
\tau_{\text{mode}}^* = \arg\max_{\tau_{\text{mode}} \geq 0} y(\tau_{\text{mode}}) \quad \text{s.t.} \quad z(\tau_{\text{mode}}) \leq Z_{\text{available}}
\]
where $Z_{\text{available}}$ is defined as the amount of available physical memory dynamically reported by workers.
This design enables Shuffle Agents to utilize memory resources on the machine that are not fully exploited by workers for in-memory shuffle.


In practice, the historical curve $(\tau_{\text{mode}}, y, z)$ is updated periodically (e.g., hourly) by an offline module, while the optimal threshold $\tau_{\text{mode}}^*$ is dynamically computed by a background daemon based on real-time $Z_{\text{available}}$ (e.g., every 30 seconds) and broadcast to Job Manager. 
Figure~\ref{fig:shuffle_mode_threshold} shows the typical range of $\tau_{\text{mode}}^*$ observed in production. 
This mechanism, deployed in Alibaba’s clusters, enables better use of idle memory for in-memory shuffle, improving memory utilization and accelerating shuffle efficiently.

\begin{figure}[t]
  \centering
  \includegraphics[width=0.9\columnwidth]{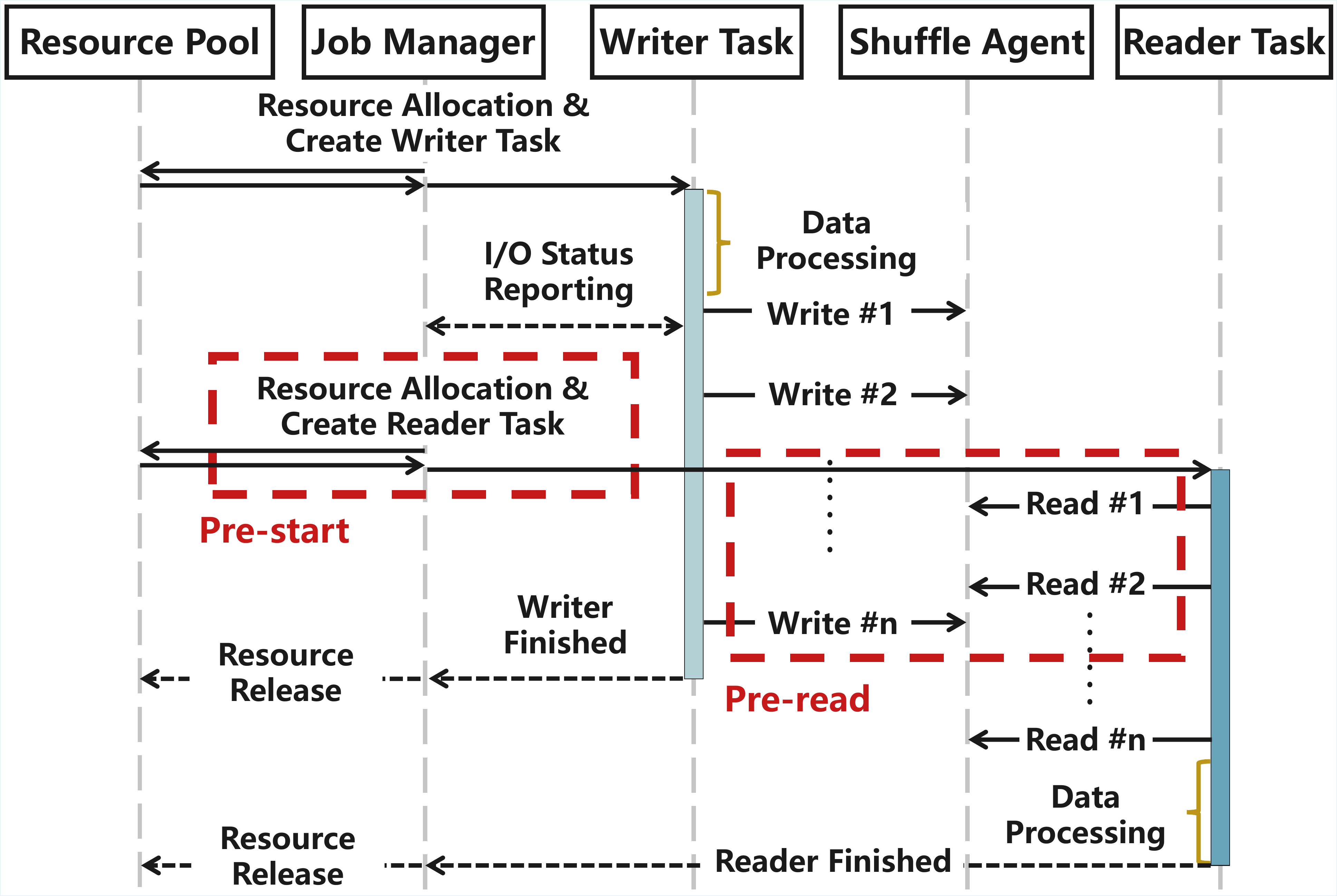}
  \trim
  \caption{The sequence diagram of progressive scheduling.}
  \trim
  \label{fig:progressive_scheduling}
\end{figure}

\subsection{Adaptive Data Reading}
\label{sec:data_fetching}

To avoid resource idleness and maximize the parallelism between upstream and downstream workers, FuxiShuffle introduces progressive scheduling, which allows the Job Manager to acquire resources on a per-worker basis rather than requesting all resources at once. This approach effectively mitigates the idle pre-launched compute nodes inherent in gang scheduling and removes the scale constraints associated with bulk resource requests. Compared with staged scheduling, it enables downstream workers to start reading shuffle data and processing earlier, improving end-to-end performance.
Figure~\ref{fig:progressive_scheduling} illustrates the shuffle workflow under progressive scheduling. The decoupling enabled by this mechanism is reflected in two key phases: Pre-start and Pre-read.



\stitle{Pre-start}
To enable pipelined parallelism between upstream and downstream tasks, progressive scheduling introduces Pre-start. During execution, upstream operators track the amount of shuffle data produced, estimate progress using historical runtime statistics, and periodically report it to the Job Manager via heartbeats.
The Job Manager aggregates progress from all upstream tasks and maintains an adaptive threshold~$\lambda_S \in [0, 1]$ for each downstream stage~$S$. A reader for stage~$S$ is scheduled \emph{only when the progress of upstream tasks reaches or exceeds}~$\lambda_S$. A lower threshold reduces end-to-end latency but increases resource consumption; a larger threshold improves resource efficiency at the cost of higher latency. The value of~$\lambda_S$ is dynamically determined based on the jobs characteristics.
In two cases, FuxiShuffle adaptively disables progressive scheduling.

\squishlist
    \item \textit{Low parallelism}: If the concurrency of task is below a threshold (e.g., <10) and requesting all resources at once is feasible, progressive scheduling is disabled in favor of gang scheduling to minimize latency for small jobs.
    
    \item \textit{Blocking downstream dependencies}: If the input edge to stage~$S$ involves a barrier (e.g., sorting, global aggregation), $\lambda_S$ is set to~1.0 to prevent reader idling and ensure complete data reading.    
\squishend

This mechanism achieves an adaptive trade-off between latency and resource utilization, ensuring that Pre-start is enabled only when the performance benefit is clear.

\stitle{Pre-read}
To break the barrier in staged scheduling where downstream readers must wait for all writers to finish, progressive scheduling introduces a Pre-read mechanism that overlaps shuffle data reading with upstream computation. Together with Pre-start, it forms a progress-aware, event-driven pipelined execution model.
The core workflow is as follows: whenever a writer commits a block of shuffle data to the Shuffle Agent, the block is immediately marked as readable. A Pre-read Handler deployed on the reader side periodically polls the Shuffle Agent to detect newly committed data. Once a reader is proactively launched through Pre-start, the Pre-read mechanism becomes effective, driving the reader to incrementally and sequentially consume committed data blocks. Instead of waiting for all shuffle data to be fully produced, the reader dynamically resolves the locations of currently available data using metadata files and reads data as soon as it becomes readable.

\stitle{Support for Skew Data Handling}
In distributed data processing, data skew often causes high job latency and low resource utilization. FuxiShuffle uses distributed Shuffle Agents with progressive scheduling to decouple upstream and downstream scheduling, enabling real-time, partition-level statistics collection (data size, record count, distinct keys). This supports adaptive skew handling, including \textit{Dynamic Partition Insertion} and \textit{Adaptive Skew-Join}~\cite{chen2021fangorn}.

\squishlist

\item \textit{Dynamic Partition Insertion} leverages FuxiShuffle’s runtime statistics to sense partition data distribution, splitting skewed ones and merging into balanced partitions. This prevents fragmented outputs and uneven task loads. After downstream processing, results are written back to their respective partition directories, ensuring data from the same partition is properly aggregated.

\item \textit{Adaptive Skew-Join} detects skew across multi-input tables using FuxiShuffle’s statistics, splitting skewed partitions of the left table and distributing them across multiple parallel join tasks, while broadcasting corresponding right-table partitions to each task to ensure efficient join execution.

\squishend

Beyond prior systems \cite{guo2016ishuffle, shen2020magnet, cheng2020ops, rao2012sailfish, wang2011hadoop, fu2018efficient} that only decouple Map phase and shuffle process, FuxiShuffle’s progressive scheduling further separates the scheduling of upstream and downstream workers and provides runtime partition-level statistics to thereby support these adaptive optimizations.

\subsection{Adaptive Data Layout Planing}
\label{sec:data-aware_layout}

\stitle{Layout Strategies}
In traditional shuffle systems, shuffle data layout is decided statically before jobs launch and cannot adapt to changing cluster conditions. 
However, production shuffle traffic is highly heterogeneous—heavy jobs prioritize stability while lightweight tasks focus on low latency. Such diversity makes static policies insufficient to balance performance, stability, and resource efficiency. FuxiShuffle addresses this by performing runtime, data-aware layout planning for each shuffle data chunk, dynamically choosing the best format and whether to enable fault tolerance.

FuxiShuffle dynamically employs multiple data layout strategies, classified into four file types in system:

\squishlist

\item \textit{Shuffle Agent File}: This is the primary storage format for the majority of shuffle data. It is generated after shuffle data is received and aggregated by the Shuffle Agent, effectively eliminating fragmented I/O read bottlenecks on the reader side.

\item \textit{Default Backup}: This refers to the backup files created when replicating shuffle data. These files are stored locally on the writer node and can be used by the reader to reconstruct data directly from the writer backup if reading the Shuffle Agent File fails, thus avoiding the need to re-execute upstream tasks.

\item \textit{Remote Backup}: Activated when the writer and Shuffle Agent reside on the same machine. In this case, backups are stored on a remote node to avoid simultaneous loss of the Shuffle Agent File and the Default Backup in a single point of failure.

\item \textit{Backup Only}: Used when the shuffle data generated by the writer is particularly large. In this mode, data is written only to the local storage of the writer, bypassing the Shuffle Agent’s merging process to conserve network bandwidth. Readers can then read data directly from the backup.

\squishend

\stitle{Layout Planning}
FuxiShuffle does not enable backups indiscriminately. Systems such as Riffle \cite{zhang2018riffle} and Magnet \cite{shen2020magnet} apply full backups for all shuffle data, which improves robustness but also brings drawbacks: longer upstream execution, and substantial I/O and storage overhead under high concurrency. In particular, writing many small backup files competes with normal shuffle I/O for disk bandwidth, often offsetting the intended fault-tolerance benefits.

To address this, FuxiShuffle introduces an adaptive backup strategy. 
During execution, the system continuously monitors writer’s runtime and partition data size, dynamically deciding whether to generate a backup and in what form, based on adaptive thresholds:

\squishlist
    \item \textit{Runtime threshold $\alpha$} (typically 5 minutes): reflects the scheduling and recomputation cost of re-execution. If a writer's execution time exceeds $\alpha$, it is classified as a heavyweight task, and a backup is generated even for small data volumes to avoid costly retries.
    
    \item \textit{Backup threshold $\beta$} (typically 500~MB): captures the cost of data retransmission. 
    If the total size of writer partition exceeds $\beta$, a backup should be enabled for the tasks, preventing them from re-execution triggered by read error.
    
    \item \textit{Backup Only threshold $\gamma$} (typically 50~MB): 
    During the shuffle process, each writer produces a corresponding amount of shuffle data for every downstream reader. If the size of a shuffle data block exceeds $\gamma$, it is considered unlikely to cause fragmented reads. In this case, the system bypasses the Shuffle Agent and writes the data directly to dual-replica backups (i.e., Backup-Only mode). This prevents large shuffle data from consuming excessive network bandwidth and I/O resources, improving overall performance while maintaining system stability.
    
\squishend

Creating backups for shuffle data is a common fault tolerance strategy in many shuffle systems, fundamentally trading redundant storage for reliable shuffle execution. FuxiShuffle’s adaptive shuffle layout planning minimizes the storage and I/O overhead while still leveraging backups for effective fault tolerance.

%% file: chapters/5-Resilience.tex
\section{Resilient Execution and Efficient Recovery}
\label{sec:recovery}

To enhance system stability, FuxiShuffle adopts proactive fault tolerance mechanisms.
First, Shuffle Agents and writers are grouped to reduce single point network pressure, each Shuffle Agent is replicated to enable failover.
Second, shuffle data availability is ensured through memory management and multi-source dynamic reading, allowing readers to reliably access consistent data.
Third, incremental error recovery enables readers to continue processing while writers are re-executed, avoiding progress loss and reducing recovery latency.
Together, these mechanisms provide highly reliable shuffle data transmission at ultra-large scale.

\begin{figure}[t]
  \centering
  \includegraphics[width=0.8\columnwidth]{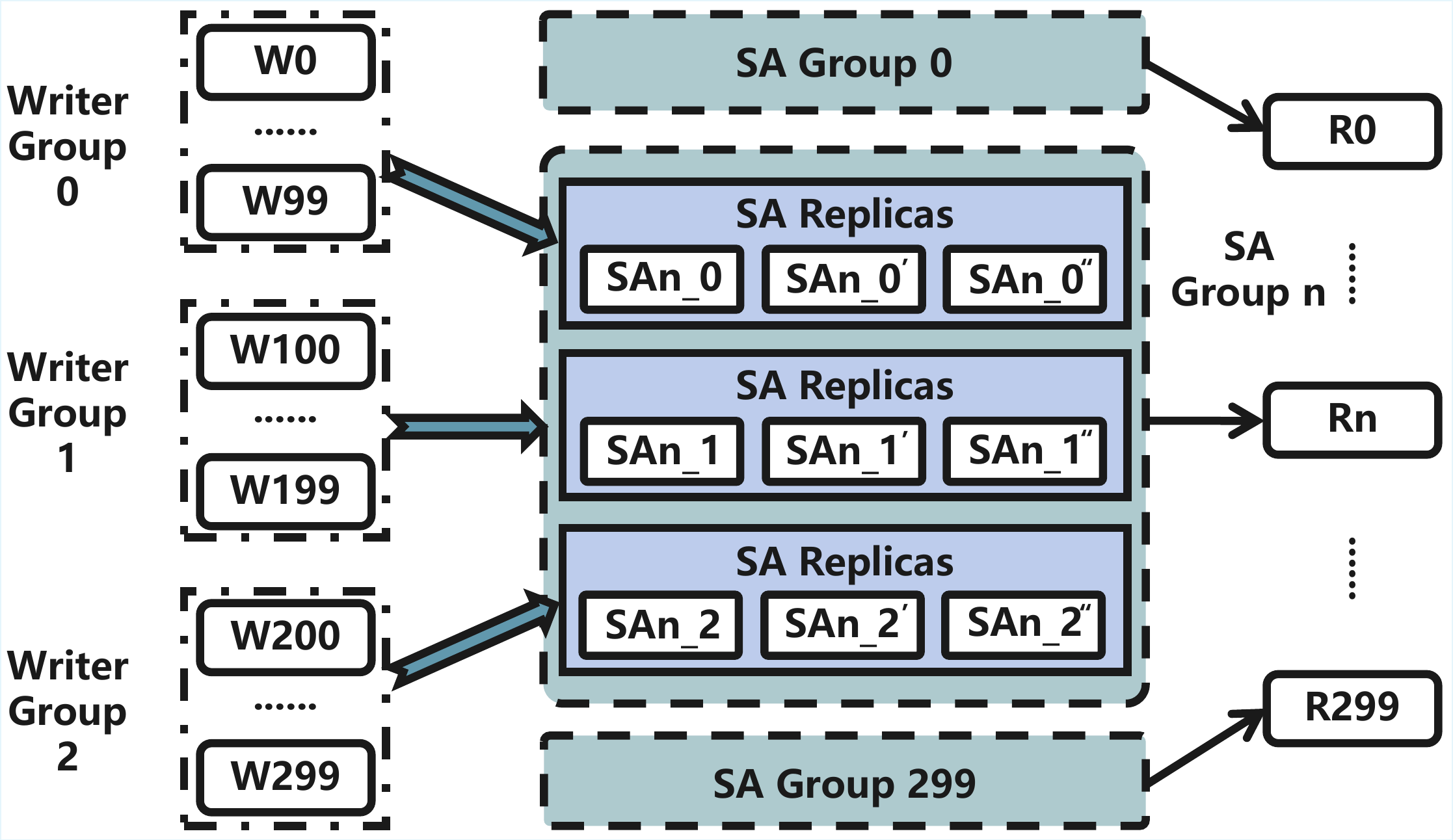}
  \trim
  \caption{Shuffle Agent grouping. The Writers and Shuffle Agents are organized into groups to avoid overloading individual Shuffle Agents, and Writers can switch between the Shuffle Agents in a shuffle group upon failures.}
  \trim
  \label{fig:shuffle_agent_group}
\end{figure}

\subsection{Shuffle Agent Grouping}


In FuxiShuffle, the Shuffle Agent is critical to the entire shuffle service, making its stability and proactive fault tolerance important. To achieve this, we introduce the concept of Shuffle Agent Grouping, which replaces the original one-to-one reader–agent mapping with a flexible many-to-many aggregation architecture.
As shown in Figure \ref{fig:shuffle_agent_group}, FuxiShuffle logically partitions upstream writers into multiple groups (e.g., 100 writers per group), with each writer group assigned to a Shuffle Agent. Therefore, multiple Shuffle Agents form a Shuffle Agent Group to collect shuffle data for a reader, where each Shuffle Agent is responsible for aggregating data from only one writer group, effectively avoiding the potential instability caused by many-to-one network fan-in in ultra-large-scale deployments. 
We introduce replication within each Shuffle Agent Group (e.g., two logical replicas per Agent). For each partition, a writer sends data to one active Shuffle Agent in the group; if the transfer fails due to network issues, memory pressure, disk overload, or machine failure, it can immediately switch to another replica and continue transmission. The replicas are implemented logically rather than as separate physical instances, thereby avoiding resource waste while improving Shuffle Agent availability and utilization.

\stitle{Data Consumption}
During data consumption, the reader does not rely on a single data source; instead, it aggregates all received data fragments from the entire Shuffle Agent Group. Based on a unified metadata management, the reader can automatically identify and merge data belonging to the same partition from multiple Shuffle Agents, ensuring result completeness and consistency. This design fundamentally addresses two key bottlenecks of traditional single-Agent architectures: (1) by partitioning upstream writers into groups, it distributes the incast pressure from ``$N$ writers $\rightarrow$ 1 Agent'' to ``$N/k \rightarrow 1$'', significantly improving network throughput and transfer success rates; and (2) by deploying multiple logical Shuffle Agents replica per group and enabling dynamic failover, it eliminates single-point-of-failure risks that could break the aggregation pipeline, thereby achieving on-disk shuffle level fault tolerance without sacrificing high performance.

\begin{figure}[t]
  \centering
  \includegraphics[width=0.65\columnwidth]{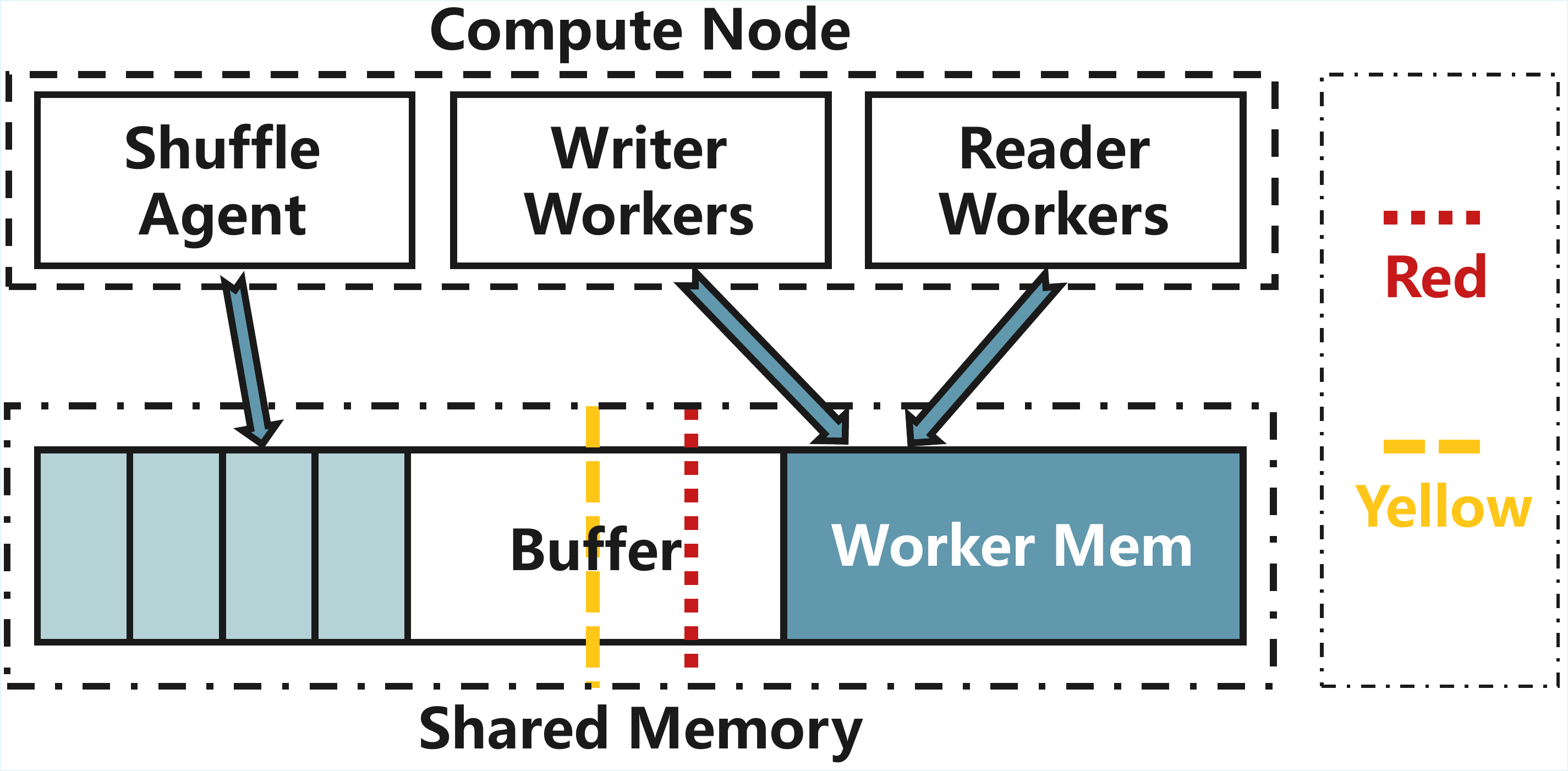}
  \trim
  \caption{Memory management for Shuffle Agents co-locating with workers on compute nodes. Smooth spilling is used when shuffle data exceeds the yellow line, while low-priority data is discarded when exceeding the red line.}
  \trim
  \label{fig:memory_assurance}
\end{figure}

\subsection{Enhancing Shuffle Data Availability}\label{subsec:memory management}

\stitle{Memory Management} In FuxiShuffle, each compute node hosts user workers (writers/readers) and a Shuffle Agent, requiring the Shuffle Agent to manage memory efficiently within a shared environment. To ensure stable in-memory shuffle, we design a dynamic memory management mechanism based on the insight that OOM is a gradual process that allows staged intervention. By defining multiple thresholds in unused memory resource, the system applies tiered, real-time policies under varying memory pressure, preventing OOM while maximizing memory utilization.


As shown in Figure \ref{fig:memory_assurance}, compute node memory is shared between the Shuffle Agent and workers. In FuxiShuffle, each shuffle write carries a priority tag, allowing the Agent to manage memory via a dual-threshold mechanism: exceeding the yellow line (e.g., 80\%) triggers smooth spilling, while crossing the red line (e.g., 90\%) removes the lowest-priority data to maintain stability. When memory pressure rises and the thresholds drop below current usage, the Agent instantly evicts data, either discarding it or spilling to disk based on urgency. Memory water levels may fluctuate in two scenarios.

\squishlist
\item \textit{Worker memory usage surging.}
The first case arises when a surge in worker memory usage pushes the Shuffle Agent’s memory usage exceeding the yellow or even red threshold. Normally, shuffle data stays below the yellow line. When available memory suddenly shrinks, shuffle memory may temporarily exceed the lines. The system then immediately evicts low-priority shuffle data to fall below the red line, and subsequently swaps additional low-priority data to disk until usage drops beneath the yellow line, restoring a stable state.

\item \textit{New shuffle data incoming.}
The second case occurs when new shuffle data are written into memory. The system reacts based on the current water level: if thresholds are not exceeded, the data are written directly; if usage crosses the yellow line, it compares priorities and swaps lower-priority data to disk; if the write would exceed the red line, lower-priority data are discarded to avoid further memory pressure.

\squishend

Memory management ensures stable in-memory shuffle. Beyond yellow/red line controls, FuxiShuffle adds dynamic fallback: heavy shuffle data triggers asynchronous disk backup, and backed-up data can be evicted under memory pressure. This provides near disk-level reliability despite memory’s non-persistence.

\stitle{Multi-source Dynamic Reading of Shuffle Data}
FuxiShuffle’s fault tolerance is rooted in its data layout planning mechanism that adapts to runtime (cf. Section \ref{sec:data-aware_layout}). This mechanism dynamically generates multiple physical replicas for each shuffle data chunk, typically including a high-performance Shuffle Agent file (in memory or on disk) and one or more Backup replicas stored on distinct nodes.
Dynamic multi-source reading of shuffle data shifts fault tolerance from passive full re-execution to proactive multi-source fallback, allowing FuxiShuffle to achieve memory-level read performance while maintaining disk-level reliability.

During data consumption, the reader employs a multi-source dynamic reading strategy: it first reads from the lowest-latency path (e.g., Shuffle Agent data in memory); upon encountering read failures or performance degradation (e.g., network timeout or unresponsive Shuffle Agent), it immediately switches to a secondary yet reliable backup path (cf. Section \ref{sec:data-aware_layout}). The specific switching behaviors of readers are as follows:

\squishlist

\item \textit{If only the Shuffle Agent fails}, the reader falls back to the complete Backup replica;

\item \textit{If only the node hosting the Shuffle Backup crashes}, the reader continues reading from the available Shuffle Agent data;

\item \textit{Even in the extremely rare event that both copies are lost simultaneously}, the system triggers only a partial upstream task re-execution to restore Backup availability.

\squishend

This co-design of adaptive layout at write time and intelligent switching at read time remains fully transparent to upper-layer computation logic, 
enabling FuxiShuffle to retrieve shuffle data from multiple data resources and thus gracefully handle diverse failure scenarios.


\subsection{Incremental Failure Recovery}

In ultra-large-scale data processing, transient hardware or network glitches often make shuffle data unavailable. Early Fuxi’s framework \cite{zhang2014fuxi} uses partial re-execution (PR): a downstream read error triggers upstream tasks re-execution to reconstruct lost data. However, like Hadoop and Spark, PR requires terminating the downstream task and waiting for upstream re-execution completed, discarding all progress. In large-scale jobs, this can cause \textit{repeated re-execution loops}, significantly increasing latency and even causing job failures.

\begin{figure}[t]
  \centering
  \includegraphics[width=\columnwidth]{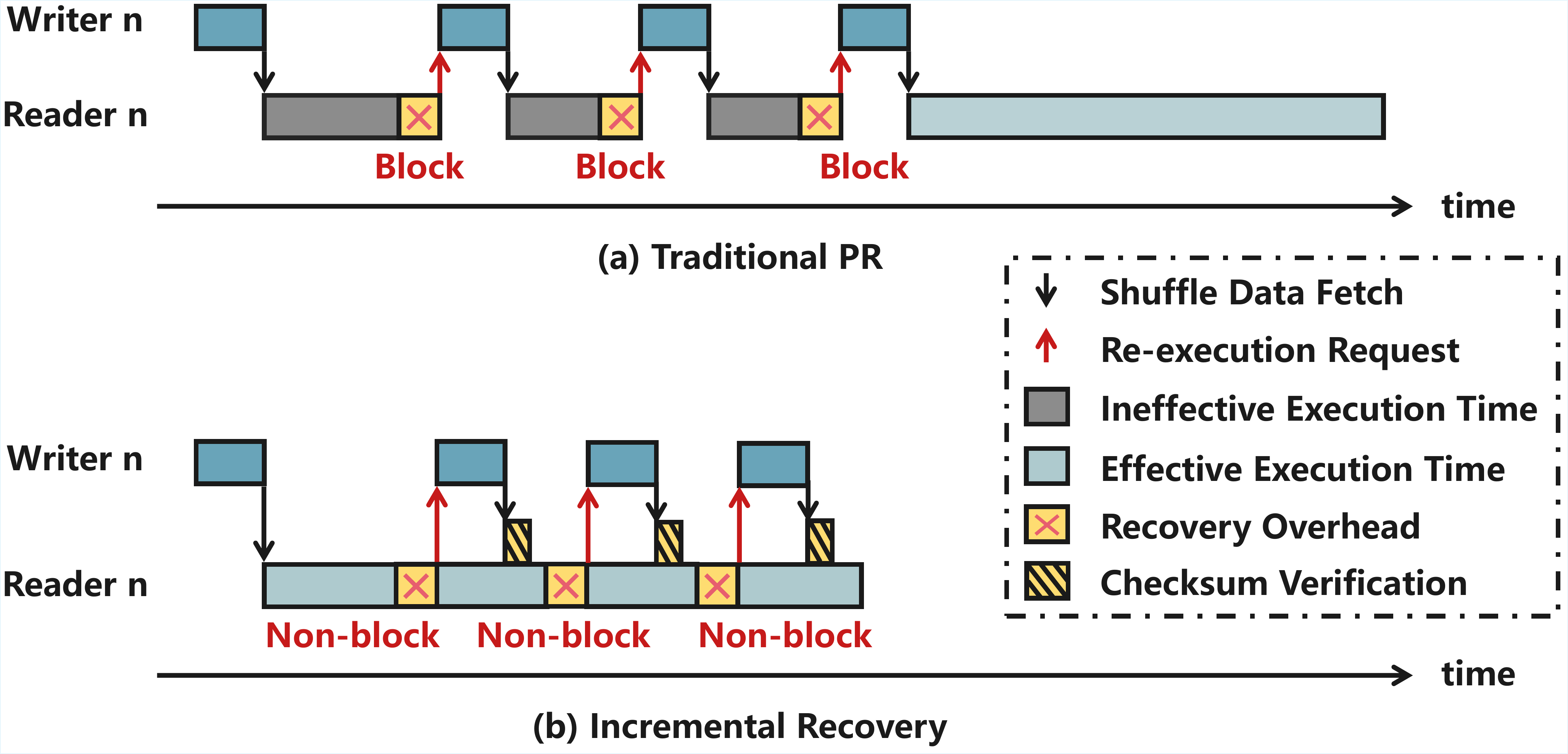}
  \trim
  \caption{Existing partial re-execution (PR) and our incremental recovery when handling repeated failures.}
  \trim
  \label{fig:incre_PR}
\end{figure}

\stitle{Incremental Recovery} FuxiShuffle introduces Incremental Recovery to overcome the serialization bottleneck of traditional partial re-execution. Its key idea is simple: when a reader fails to fetch a subset of shuffle data, it continues processing the valid data while the system re-executes the corresponding writer to regenerate the missing data. Once the missing data is regenerated, the reader incrementally loads only that portion.
This converts the classic serial recovery model into a parallel one, allowing downstream computation and upstream re-execution to proceed concurrently, significantly reducing end-to-end latency and effectively containing the blast radius of error propagation.
As shown in Figure \ref{fig:incre_PR}, traditional PR forces readers to block and discard progress, whereas Incremental Recovery keeps the reader running and performs checksum verification upon receiving regenerated data.
In Alibaba MaxCompute, this mechanism lowers both error and retry count per job, largely due to the reduced completion time for failed jobs.


\stitle{Analysis of Correctness}
Incremental Recovery depends on a checksum-driven version consistency mechanism to ensure correctness, as mismatched versions or partitioning schemes between old and new data would otherwise produce incorrect shuffle results.
Before resumption, the reader fetches each writer’s checksum and version (RetryIdx, Section \ref{sec:metadata_management}) from the Job Manager and proceeds only if missing data is locatable and remaining data is valid; otherwise, a full re-execution is triggered. During recovery, re-executed writers produce new-version data with checksums, which the reader selectively pulls. After loading, the reader verifies the aggregated checksum against the Job Manager (Figure \ref{fig:incre_PR}); any missing or duplicate data aborts the process. This dual-phase verification enables strong consistency and semantic equivalence to full re-execution.

%% file: chapters/6-Evaluation.tex
\section{Evaluation}

To comprehensively evaluate the performance of FuxiShuffle, we conduct three sets of experiments on Alibaba’s internal test clusters: (1) a comparative study against existing systems; (2) an evaluation of the adaptability of FuxiShuffle; and (3) a validation of the fault tolerance of FuxiShuffle under failure scenarios.

\subsection{Experiment Settings}

\stitle{Hardware Environment}
All experiments are conducted on an Alibaba internal test cluster, which is a computing platform with the following configuration:  

\squishlist

\item \textit{Compute nodes:}
20 machines, each of which is an \texttt{x86\_64} Linux server equipped with 96 logical CPU cores (Intel Xeon Platinum 8163 CPU $@$ 2.50 GHz) and 412 GB RAM.

\item \textit{Storage nodes:} 18 machines, each configured with 13 $\times$ 8 TB HDDs ($\approx$104 TB raw storage per node, totaling $\approx$1.87 PB), dedicated to storing intermediate shuffle data and persistent outputs.  

\squishend


\stitle{Performance Metrics}
We adopt the following main metrics to evaluate system performance in our experiments.

\squishlist

\item \textit{End-to-End Runtime} (E2E Runtime, s): The total elapsed time from job submission to completion, measuring the time performance.

\item \textit{Scheduled Resource Consumption} (CU Cost, Core·min): 
It quantifies the total resource capacity reserved by the Fuxi scheduler for a job, calculated as the sum of (allocated resource specification $\times$ scheduling occupancy time) across all tasks. 

\squishend

\stitle{Workloads}
To evaluate system performance, we select two widely adopted benchmark workloads: \textit{TeraSort} \cite{o2009winning} and \textit{TPC-DS} \cite{poess2007you}.

\squishlist

\item \textit{TeraSort} is a classic sorting-intensive benchmark. It is highly sensitive to critical factors like network bandwidth, disk I/O, memory management, and fault tolerance, and is specifically designed to evaluate the throughput and scalability of distributed systems during large-scale shuffle phases.

\item \textit{TPC-DS} defines a rich set of complex SQL queries involving multi-table JOINs, subqueries, window functions, aggregations, and sorting—covering typical OLAP operations. It incorporates real-world challenges such as data skew and sparse joins, making it highly effective for testing system performance and robustness.

\squishend

 
In \textit{TPC-DS}, ~60\% of queries involve multi-table JOINs, generating substantial shuffle traffic and reflecting mixed production workloads, making it suitable for overall performance evaluation. \textit{TeraSort} is shuffle-intensive, highlighting both overall system and shuffle performance. Both benchmarks are used in our experiments, with scales adjusted per test.

\subsection{Overall Performance Evaluation}


\stitle{Baselines}
To rigorously assess the performance and practicality of FuxiShuffle, we compare against two representative production-grade shuffle implementations—covering both MapReduce-style and DAG-based execution paradigms:

\squishlist

\item \textit{Hadoop-like}: An \textit{deeply} optimized and customized variant of Apache Hadoop MapReduce’s shuffle architecture \cite{apache_hadoop}, deployed and battle-tested in Alibaba’s MaxCompute for years.

\item \textit{Spark-like}: An \textit{basic} optimized version of Apache Spark \cite{apache_spark}, also deployed on scale in Alibaba Maxcompute production clusters.

\item \textit{FuxiShuffle On-disk} (Ours): Our implementation configured to use only the on-disk shuffle mode.

\item \textit{FuxiShuffle} (Ours): Our implementation utilizing in-memory shuffle mode and on-disk shuffle mode adaptively where beneficial. 

\squishend

\begin{figure}[t]
  \centering
  \includegraphics[width=\linewidth]{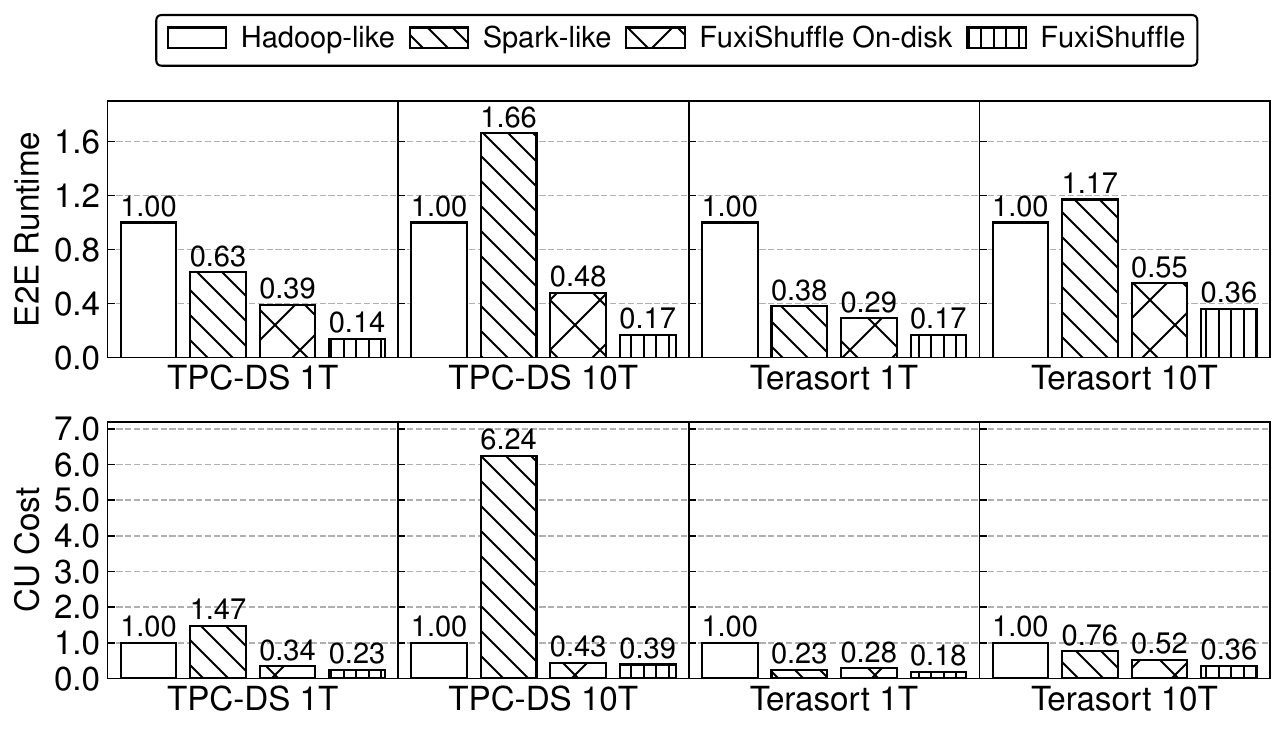}
  \trim
  \caption{E2E runtime and CU cost of FuxiShuffle and baseline systems for TPC-DS and TeraSort on 1TB and 10TB scale.}
  \trim
  \label{fig:preformance_evaluation}
\end{figure}

\stitle{Result Analysis}
We normalize the performance metrics of other systems to relative ratios, that is, speedup or overhead factors, using the performance of the Hadoop-like system across various workloads and data scales as the baseline, enabling fair and unified cross-system and cross-workload comparison. The results are shown in Figure \ref{fig:preformance_evaluation}.
Across varying workloads and data scales, FuxiShuffle consistently achieves the best performance, while FuxiShuffle on-disk delivers the second-best or near-second-best performance in all scenarios. This advantage comes from FuxiShuffle’s Shuffle Agent centric distributed architecture, which collects and aggregates shuffle data, thereby effectively mitigating the I/O and network bottlenecks caused by fragmented reads and massive concurrent connections in traditional approaches. Additional optimizations, including the \textit{Adaptive Shuffle Mode Selection}, \textit{Adaptive Data Layout Planning} and the \textit{Adaptive Data Reading} mechanism, further contribute to FuxiShuffle’s performance superiority.
Moreover, the results confirm that FuxiShuffle consistently outperforms its on-disk mode, as it prioritizes memory as the shuffle medium and accurately identify the shuffle traffic suitable for in-memory shuffle. This minimizes disk I/O and enables faster read/write operations, thereby significantly improving shuffle efficiency.


Under certain workloads, we observed that the Spark-like system did not outperform Hadoop-like, as it was less optimized and, with 10 TB workloads, insufficient memory forced Spark-like system to frequently spill partitions to disk, further reducing its efficiency.

\subsection{Adaptability Evaluation}

\stitle{Adaptive Shuffle Mode Selection}
To evaluate the effectiveness of the \textit{Shuffle Mode Selection} mechanism, we design two types of workloads: CPU-intensive jobs (low shuffle ratio, job size 100 × 100, five jobs, submitted in batches) and I/O-intensive jobs (high shuffle ratio, job size 50 × 50, ten batch-submitted jobs $\&$ ninety JIT-submitted jobs to maintain high cluster load). We simulate real-world scenarios by adjusting the computation-to-I/O ratio.
Given that \textit{Shuffle Mode Selection} is tightly coupled with \textit{Memory Management} machinism (Section \ref{subsec:memory management}), we conduct ablation experiments (enabling or disabling them in different combinations) to systematically analyze their respective contributions to overall performance.

\begin{table}[]
\renewcommand{\arraystretch}{1}
\caption{The impact of \textit{Shuffle Mode Selection} (SMS) and \textit{Memory Management} (MM). E2E Runtime, Write Time, and Read Time are measured in seconds (s), while CU Cost is in Core·minutes (Core*min), other tables adopt the same units.}
\trim
\label{tab:sms_mm_ablation}
\resizebox{\linewidth}{!}{%
\setlength{\tabcolsep}{2pt}
\begin{tabular}{c cccc}
\hline
\rowcolor[HTML]{F7F8FC}
\textbf{} &
  \textbf{w/o SMS\&MM} &
  \textbf{w/o MM} &
  \textbf{w/o SMS} &
  \textbf{FuxiShuffle} \\ \hline
\textbf{E2E Runtime}         & 3,677    & 3,270    & 1,995    & 1,318    \\
\textbf{CU cost}   & 44,166   & 38,939   & 23,707   & 15,474    \\
\textbf{Write Time} & 375,168  & 403,218  & 64,689   & 54,244    \\
\textbf{Read Time}  & 1,830,526 & 1,480,058 & 888,831  & 538,739   \\
\textbf{MemoryShuffleUtil}    & 8.50\%  & 15.90\% & 52.44\% & 69.02\% \\ \hline
\end{tabular}%
}
\trim
\end{table}

\begin{figure}[t]
  \centering
  \includegraphics[width=0.75\linewidth]{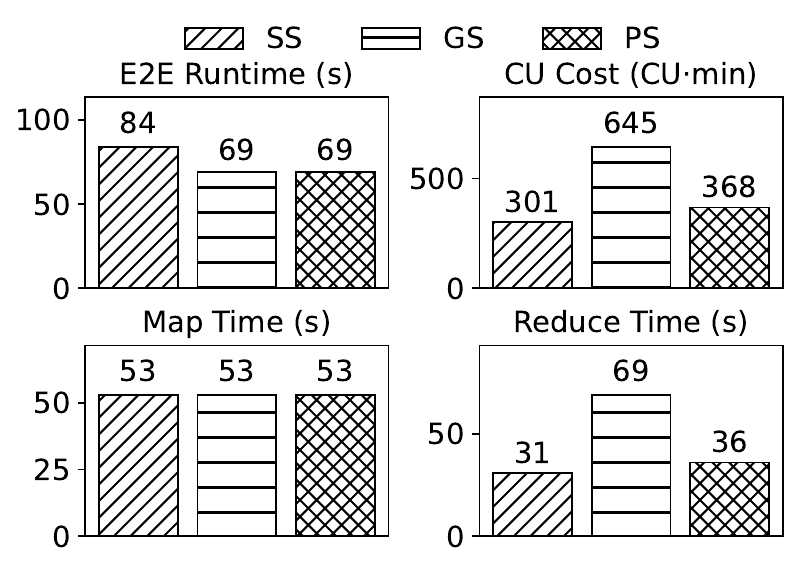}
  \trim
  \caption{The performance of FuxiShuffle when using three scheduling modes: \textit{Staged Scheduling} (SS), \textit{Gang Scheduling} (GS), and \textit{Progressive Scheduling} (PS).}
  \trim
  \label{fig:data_fetching}
\end{figure}

\begin{figure}[t]
  \centering
  \includegraphics[width=0.75\linewidth]{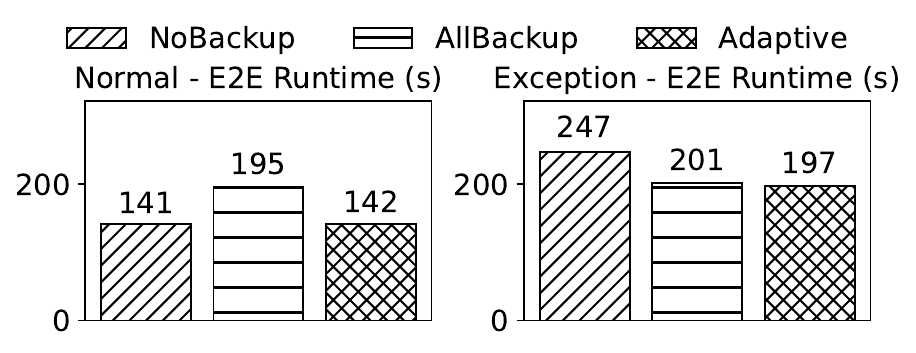}
  \trim
  \caption{The performance of FuxiShuffle when using three data backup strategies: \textit{NoBackup}, \textit{AllBackup}, and \textit{Adaptive}. \textit{Normal} does not introduce failures, while \textit{Exception} introduces a single shuffle data reading failure for each job.}
  \trim
  \label{fig:adaptive_backup}
\end{figure}

From the experimental results presented in Table \ref{tab:sms_mm_ablation}, we observe that disabling either the adaptive \textit{Shuffle Mode Selection} or the Shuffle Agent’s \textit{Memory Management} leads to significant performance degradation. 
The \textit{Memory Management} mechanism allows us to utilize more memory on a single machine, while the adaptive \textit{Shuffle Mode Selection} enables choosing the most suitable mode for each task to use memory efficiently. Combined, these two mechanisms minimize both Write Time and Read Time in this experiment, resulting in significant advantages in E2E Runtime and CU Cost.
This indicates that both the adaptive \textit{Shuffle Mode Selection} and the \textit{Memory Management} of Shuffle Agent are crucial contributors to the performance advantages of FuxiShuffle.

\stitle{Adaptive Data Reading}
To evaluate \textit{Adaptive Data Reading}, we construct test jobs with highly parallel, non-blocking downstream tasks, comparing three modes in FuxiShuffle: staged scheduling (SS) where readers start after all writers finish, gang scheduling (GS) with simultaneous writers and readers, and progressive scheduling (PS) where readers launch adaptively based on progress of writers.

The experimental results, shown in Figure \ref{fig:data_fetching}, demonstrate that under the staged scheduling, the writer and reader execute without overlap, leading to the longest E2E Runtime. However, because the readers do not start early and wait for upstream outputs, their individual runtimes are minimized, resulting in the lowest overall CU Cost among the three modes.
In gang scheduling, writers and readers start simultaneously, and readers occupy resources while waiting for the writer’s output. Although this achieves the shortest E2E Runtime, it incurs significantly higher resource consumption and costs.
Progressive scheduling allows readers to start adaptively based on the writer’s progress, achieving an E2E Runtime nearly identical to gang scheduling while reducing unnecessary waiting. As a result, it minimizes CU Cost and delivers the best balance between runtime and resource efficiency.

\stitle{Adaptive Data Layout Planning - Default Backup}
To evaluate the adaptive Default Backup mechanism in \textit{Adaptive Data Layout Planning}, we run a job with 300 map and 300 reduce tasks, including a few large (10,000 MB) and many small (1,000 MB) instances to mimic real-world workload imbalance. We compare three modes in FuxiShuffle
: NoBackup (no Default Backup), AllBackup (all shuffle data backed up), and Adaptive (only critical shuffle data backed up). Exception scenario refers to injecting one exception into the job.

As shown in Figure \ref{fig:adaptive_backup}, under the Normal scenario, the Adaptive mode identifies instances with low re-execution cost based on their data size and execution time, skipping writing Default Backup files. 
As a result, its E2E Runtime is close to that of the NoBackup, which writes no backup at all, and significantly more efficient than the AllBackup, which incurs substantial time overhead during backup preparation.
Under the Exception scenario, the Adaptive mode even achieves the best E2E Runtime. It has already created backups for instances with high re-execution cost and therefore only needs to rerun upstream instances with low re-execution cost. In contrast, NoBackup must rerun all instances affected by the exception.
In summary, the Adaptive mode achieves performance close to NoBackup in the Normal scenario, while offering stability in the Exception scenario comparable to that of the AllBackup mode.


\begin{table}[]
\caption{Disabling Backup Only (BO) on FuxiShuffle.}
\trim
\label{tab:Backup_Only}
\renewcommand{\arraystretch}{} 
\setlength{\tabcolsep}{16pt}
\resizebox{\linewidth}{!}{
\begin{tabular}{
>{\columncolor[HTML]{FFFFFF}}c cc}
\hline
\rowcolor[HTML]{F7F8FC}
\textbf{} & \textbf{w/o BO} & \textbf{FuxiShuffle} \\ \hline
\textbf{E2E Runtime}         & 534   & 395  \\
\textbf{Writer Task Runtime}      & 229   & 100  \\
\textbf{Reader Task Runtime}   & 305   & 295 \\
\textbf{CU Cost}   & 327   & 144  \\
\textbf{Write Time} & 15,120 & 3,879 \\
\textbf{Read Time}  & 811   & 796  \\ \hline
\end{tabular}
}
\trim
\end{table}


\stitle{Adaptive Data Layout Planning - Backup Only}
To investigate the adaptivity of the \textit{Adaptive Data Layout Planning} mechanism in creating Backup Only for large-size shuffle partition, we construct a job with significant data skew: a total data volume of 400GB, job size 100(map) × 100(reduce). Among them, the skewed partition holds 360GB (90\% of the total), while the remaining partitions evenly share the remaining 40GB.
We configure FuxiShuffle in the on-disk mode and set the threshold for the Backup Only policy such that when the data size of a single shuffle partition exceeds 1GB, only a Backup Only file is generated, skipping transmission to the Shuffle Agent. 
The experiment evaluates the ability of \textit{Adaptive Data Layout Planning} in avoiding large-partition transfer overhead comparing performance metrics with the Backup Only enabled and disabled.


As shown in Table \ref{tab:Backup_Only}, enabling Backup Only eliminates the need to transfer large shuffle data over the network to Shuffle Agents, significantly reducing shuffle Write Time. With fewer requests for Shuffle Agents to handle, data from other partitions can be processed faster, thereby shortening the E2E Runtime and reducing CU Cost. Moreover, for the cluster, large shuffle data only needs to exist as Backup Only files, alleviating I/O pressure on cluster storage and allowing small shuffle data to be handled more quickly. Overall, Backup Only efficiently simplifies the transfer of large shuffle data, improving the overall performance of FuxiShuffle. This demonstrates the dual performance and cost benefits of the \textit{Adaptive Data Layout Planning} mechanism.

\begin{table}[]
\caption{The performance of FuxiShuffle under Read and Write failures. \textit{No Fault} means normal execution without failures, and \textit{w/o FT} disables fault tolerance.}
\trim
\label{tab:Targeted_Fault_injection}
\resizebox{\linewidth}{!}{
\setlength{\tabcolsep}{1.5pt}
\begin{tabular}{cccccc}
\hline
\rowcolor[HTML]{F7F8FC}
  &
  &
  \multicolumn{2}{c}{\textbf{Write Fault}} &
  \multicolumn{2}{c}{\textbf{Read Fault}} \\ \cline{3-6}
  \rowcolor[HTML]{F7F8FC}
\multirow{-2}{*}{\textbf{}} &
  \multirow{-2}{*}{\textbf{No Fault}} &
  \textbf{FuxiShuffle} &
  \textbf{w/o FT} &
  \textbf{FuxiShuffle} &
  \textbf{w/o FT} \\ \hline
\textbf{E2E Runtime}         & 681   & 755   & 873   & 703   & 1,198  \\
\textbf{CU Cost}      & 7,273 & 7,818 & 9,233 & 7,491 & 1,1643 \\
\textbf{Rerun Times}            & 0     & 27    & 721   & 0     & 1,024  \\
\hline
\end{tabular}
}
\trim
\end{table}

\begin{table}[]
\caption{The performance of FuxiShuffle under random fault injection. \textit{No Fault} means normal execution without failures, and \textit{w/o FT} disables fault tolerance.}
\label{tab:random_fault_injection}
\trim
\resizebox{\linewidth}{!}{
\setlength{\tabcolsep}{12pt}
\begin{tabular}{cccc}
\hline
\rowcolor[HTML]{F7F8FC} 
   &
   &
   \multicolumn{2}{c}{\textbf{Random Fault}} \\ \cline{3-4} 
  \rowcolor[HTML]{F7F8FC}
\multirow{-2}{*}{} &
  \multirow{-2}{*}{\textbf{No Fault}} &
  \textbf{FuxiShuffle} &
  \textbf{w/o FT} \\ \hline
\textbf{E2E Runtime}       & 5,636 & 6,177  & 15,127 \\
\textbf{CU Cost} &
  74,338 &
  80,365 &
  {142,776} \\
\textbf{Rerun Times}       & 0     & 9      & 28,138 \\
\hline
\end{tabular}
}
\trim
\end{table}

\subsection{Resilience Evaluation}


\stitle{Fixed Fault Injection}
In fixed fault-injection experiments, we run two job types to simulate heterogeneous workloads: large jobs (512 × 512, 3 GB input per writer task) and small jobs (256 × 256, 300 MB input per writer task), with controlled delays to mimic real scenarios. Each round submits one large and two small jobs concurrently to test fault tolerance under mixed workloads. Faults are injected by disconnecting a compute or storage node for 30 seconds in different phases. Specifically, there are two types:

\squishlist

\item \textit{Write Fault Injection}: disconnect the network during the Shuffle Write phase to simulate a failure on the writing side.

\item \textit{Read Fault Injection}: disconnect the network during the Shuffle Read phase to simulate a failure on the reading side.

\squishend
 
We evaluate FuxiShuffle’s performance under both configurations with the proactive fault tolerance (FT) mechanism disabled and enabled, in order to quantify the actual benefits of its capability.
As shown in Table~\ref{tab:Targeted_Fault_injection}, the experiments demonstrate that FuxiShuffle’s proactive fault tolerance mechanism can effectively withstand single point of failures. When a fault is injected during the write phase, disabling fault tolerance results in 721 task retries, significantly increasing runtime and CU Cost. 
In contrast, FuxiShuffle leverages \textit{Shuffle Agent Grouping} to enable replica-based continuation on failed nodes, allowing large tasks to complete without retries and only a minimal number of small tasks without Backup to undergo lightweight retries. During these small task retries, FuxiShuffle’s \textit{Incremental Failure Recovery} mechanism accelerates the recovery process, causing only a slight increase in the metrics.
During failures in the read phase, large tasks must trigger upstream recomputation when fault tolerance is disabled, nearly doubling runtime and CU Cost. FuxiShuffle, however, pre-places remote backups (Remote Backup and Backup Only files) for critical data through \textit{Adaptive Data Layout Planning}, ensuring that large tasks complete without retries and small tasks finish promptly. Overall, in the case of single point of failures, FuxiShuffle keeps all metric fluctuations within 10\%, demonstrating high availability and robustness.

\stitle{Random Fault Injection}
In random fault-injection experiments, we used a uniform workload of 256 × 256 tasks per job, with 1 GB input per Map task. Five jobs ran concurrently throughout, and 50 jobs were completed, allowing us to evaluate FuxiShuffle’s stability under sustained high load.
Faults were injected periodically: every 12–15 minutes, a compute or storage node was randomly disconnected for 30 seconds to emulate transient production-like network failures. We compared FuxiShuffle with proactive fault tolerance enabled versus disabled to assess the effectiveness and robustness of the mechanism under continuous disturbances.

As shown in Table~\ref{tab:random_fault_injection}, without fault injection, FuxiShuffle achieves the lowest E2E Runtime and CU Cost, with a small amount of read Backup caused by occasional I/O failures due to high disk load. After fault injection, FuxiShuffle’s E2E Runtime and CU Cost increase only by about 9.6\% and 8.1\%, respectively, based on its proactive fault tolerance mechanism. Most failures trigger only lightweight read of Backup files, and only very few tasks require retries due to consecutive I/O failures. In contrast, with fault tolerance disabled, E2E Runtime increases nearly threefold and CU Cost doubles, performing significantly worse than the version with fault tolerance enabled. This is because many tasks undergo 1 to 3 retries, and with the Backup mechanism disabled, a large number of tasks are retried, competing for I/O resources, and some tasks experience \textit{repeated re-execution loops}, forming a vicious cycle, highlighting the necessity of the proactive fault tolerance in FuxiShuffle. 

%% file: chapters/7-Lessons_in_Practice.tex
\subsection{Insights from Practical Deployment}

Besides our designs and evaluations, we also summarize a few practical insights for deploying efficient shuffle service.

\stitle{Data Locality} On compute nodes, Shuffle Agents leverage idle memory reserved by co-located workers for peak usage, enabling in-memory shuffle. Readers should be co-located with the Shuffle Agents holding their required data during scheduling to reduce network overhead. Similarly, when running Shuffle Agents on storage nodes for disk shuffle, they should co-locate with Alibaba's Pangu Storage Agents so that shuffle data can be written locally.

\stitle{Resource Guarantees} Shuffle Agents usually co-locate with other Workers on the same machine, and we ensure Shuffle Agents operate efficiently via resource protection. In particular, the Shuffle Agent process is exclusively bound to a set of CPU cores, and these dedicated cores consume only 3\%–5\% of each machine’s CPU.

\stitle{Elasticity} Workload may fluctuate violently for production clusters. Owners of the compute jobs can purchase more VMs at spikes, and we provide elasticity for shuffle service at the cluster level. When sustained high Shuffle IO pressure is detected, we provision more VMs with cloud disks and dynamically expand the cluster to provide more resources for shuffle service.

%% file: chapters/8-Related_Work.tex
\section{Related Work}\label{sec:related}

\stitle{Distributed Data Processing Systems}
Over the past two decades, distributed data processing has evolved from early batch processing frameworks to cloud-native and serverless architectures \cite{taleb2024amazon, kang2025abase, li2025adaptive, chen2025oceanus}. In particular, Google MapReduce \cite{dean2008mapreduce} and Hadoop \cite{apache_hadoop} pioneered large-scale parallel computing on commodity clusters. Spark \cite{apache_spark} introduced DAG-based execution and in-memory computation and has become a dominant open-source engine. Major companies also developed their customized systems with diverse features: Google FlumeJava \cite{chambers2010flumejava} and Dataflow \cite{krishnan2015google} introduced unified batch–stream processing; Microsoft Cosmos/Scope \cite{patel2019big, chaiken2008scope} supported massive log and search pipelines; our MaxCompute at Alibaba \cite{wang2018billion}, powered by Fuxi \cite{zhang2014fuxi, chen2021fangorn}, featured as a serverless data warehouse with efficient multi-tenancy; Tencent Oceanus \cite{chen2025oceanus} and Huawei MRS \cite{huawei_mrs} also enhanced distributed data processing from the aspects of availability, scalability, and intelligence. Since data shuffle is a major performance bottleneck for distributed data processing~\cite{shen2020magnet,guo2016ishuffle,cheng2020ops}, the designs and lessons learned from our FuxiShuffle may help improve efficiency for these systems.


\stitle{Data Shuffle Systems} 
Due to the importance of data shuffle, several systems are designed to improve its performance as introduced in Section~\ref{sec:intro}. We compare our FuxiShuffle with existing systems horizontally in Table~\ref{tab:comparison}.
Existing approaches rely on static shuffle mode, making it hard to balance performance and resource efficiency under diverse workloads.
In contrast, FuxiShuffle dynamically selects the optimal shuffle mode based on runtime conditions, maximizing in-memory acceleration under resource constraints and accurately identifying shuffle traffic suitable for memory processing.
For shuffle decoupling, FuxiShuffle uses progressive scheduling to adaptively schedule diverse operators and and parallelism levels while collecting fine-grained data and skew statistics to guide downstream decisions.
For fault tolerance, FuxiShuffle groups Shuffle Agents and writers to reduce network pressure and prevent single points of failure, dynamically manages memory, provides backup failover, and lets downstream tasks continue computing during upstream re-execution, enabling proactive recovery.
Overall, FuxiShuffle shows clear advantages in shuffle mode, decoupling, and fault tolerance.
Although these ideas are not fundamentally new, we are arguably the first to present them and perhaps more importantly their targeted problems in the context of running large-scale shuffle services at production scale.

\begin{table}[]
\caption{FuxiShuffle versus existing data shuffle systems.}
\trim
\label{tab:comparison}
\resizebox{\linewidth}{!}{%
\begin{tabular}{cccc}
\hline
\rowcolor[HTML]{F7F8FC}
\multicolumn{1}{l}{\textbf{System}} &
  \multicolumn{1}{l}{\textbf{Shuffle Mode}} &
  \multicolumn{1}{l}{\textbf{Shuffle Decoupling}} &
  \textbf{Fault Tolerance} \\ \hline
Riffle \cite{zhang2018riffle}& Disk   & Partial        & Partially Proactive \\
iShuffle \cite{guo2016ishuffle}   & Disk   & Stage-level    & Partially Proactive                              \\
Magnet \cite{shen2020magnet}     & Disk   & Stage-level & Partially Proactive                         \\
Sailfish \cite{rao2012sailfish}   & Disk   & Pipeline        & Passive                         \\
OPS \cite{cheng2020ops}        & Memory & Pipeline        & Passive                         \\
HDShuffle \cite{qiao2019hyper} & Disk   & Stage-level     & Passive              \\
Hadoop-A \cite{wang2011hadoop} & Memory & Pipeline        & Passive             \\
SCache \cite{fu2018efficient}    & Memory & Pipeline        & Passive         \\
\textbf{FuxiShuffle} &
  \textbf{Adaptive} &
  \textbf{Adaptive} &
  \textbf{Proactive} \\ \hline
\end{tabular}
}
\trim
\end{table}

%% file: chapters/9-Conclusion.tex
\section{Conclusion}

This paper presents FuxiShuffle, an adaptive and highly reliable shuffle service tailored for the ultra-large-scale production environment of Alibaba MaxCompute. FuxiShuffle leverages a distributed set of Shuffle Agents to provide adaptive shuffle service. It dynamically chooses the shuffle mode, schedules upstream and downstream workers, and sets backups per shuffle unit. For fault tolerance, FuxiShuffle groups Shuffle Agents with multi-replica failover, ensuring data robustness for in-memory and on-disk shuffle and enabling fast incremental recovery.
Evaluation on Alibaba Cloud clusters shows that FuxiShuffle significantly outperforms baseline systems, reducing average end-to-end job execution time by 76.36\% and resource consumption by 67.14\%. Under single-node failures, all metrics vary by less than 10\%, and the system maintains high efficiency and strong robustness even under continuous disturbances.